\documentclass[11pt, letterpaper, twocolumn, copyright, goog]{google}
\usepackage[authoryear, sort&compress, round]{natbib}
\usepackage{afterpage}
\usepackage{microtype}
\usepackage{placeins}
\usepackage{stfloats}
\usepackage{xurl}
\extrafloats{100}

\keywords{generative AI; image generation and editing; multimodal large language models; differential privacy; technology adoption; global diffusion; digital marketing; e-commerce; educational technology\\[0.3em]
}

\uselogo{}

\title{How are viral trends different from educational and commercial uses of LLM image generation and editing? A privacy-preserving descriptive analysis}

\correspondingauthor{tuckerw@google.com, marikaswanberg@google.com}

\reportnumber{}

\renewcommand{\today}{2026-11-4}

\author[*,1]{Will Tucker-Ray}
\author[*,2]{Marika Swanberg}
\author[*,2]{Alexander Knop}
\author[3]{Marvin Chow}
\author[1]{Eric Culbertson}
\author[4]{Julian Jacobs}
\author[2]{Pritish Kamath}
\author[2]{Ravi Kumar}
\author[5]{Jason Mothersill}
\author[6]{Mikhail Pravilov}
\author[7]{Eugene Quan}
\author[6]{Yurii Sushko}

\affil[*]{Equal contributions}
\affil[1]{Google Behavioral Economics}
\affil[2]{Google Research}
\affil[3]{Google Marketing}
\affil[4]{Google DeepMind}
\affil[5]{Google Learning}
\affil[6]{Google Privacy, Security, \& Safety}
\affil[7]{Google Gemini Data Science}

\begin{abstract}
Gemini's Nano Banana models, first introduced in late 2025, led to a shift in the capabilities and use of image generation and editing. We utilize Urania, a Google-developed framework for generating insights about LLM chatbot interactions with rigorous differential privacy (DP) guarantees, to analyze use cases and intents, techniques and ways of prompting, without ever looking at a single image or individual prompt. We select a subsample of 2.1M conversations in early 2026 which utilized Nano Banana models, into roughly 800 clusters in five categories: (1) Social \& Viral Sharing, (2) Brand \& Marketing Collateral, (3) Educational \& Technical Information Media, (4) Product Visualizations \& E-commerce, and (5) Architectural \& Spatial Designs. We then compare these clusters on 21 features of use, measuring aspects of prompts' (a) editing techniques, (b) artistic style requests, (c) composition, and (d) chatbot interactions. This allows us to describe a baseline of how people are using image generation and editing tools for different use cases---(including growing and emergent use cases which were not possible before Nano Banana)---in a privacy-preserving manner.
\end{abstract}

\newcommand{\pct}[2]{%
  \makebox[2.6em][r]{#1\%}\hspace{5pt}%
  \makebox[44pt][l]{\textcolor{NavyBlue!45}{\rule{\fpeval{(#1)/(#2)*44}pt}{5.5pt}}}%
}

\begin{document}
\maketitle

\section{Introduction}
\label{sec:introduction}

Large language models (LLMs) are tools widely used by more than a billion people monthly, through chatbots like \href{https://chatgpt.com/}{ChatGPT}, \href{https://gemini.google.com/}{Gemini}, \href{https://claude.ai/}{Claude}, \href{https://chat.deepseek.com/}{DeepSeek}, and \href{https://google.com/aimode}{Google AI Mode}. One of the uses people bring to these tools is the generation of new images or modification or editing of an existing image, made possible by advances in LLMs to process multimedia data \citep{raeini2025}. In late 2025, Google \href{https://blog.google/products-and-platforms/products/gemini/updated-image-editing-model/}{released} the Nano Banana model \citep{arstechnica2025}, which allowed users to blend multiple reference images into a single image, maintain character consistency, use models' world knowledge to generate and edit images---representing a \href{https://developers.googleblog.com/en/introducing-gemini-2-5-flash-image/}{step change}\nocite{googledevs2024} in the capability of LLM-based image generation and editing.

Following the launch of Nano Banana, there was a rise in image generation. Within a week, \href{https://x.com/joshwoodward/status/1963627742618165270?s=19}{10 million people}\nocite{woodward2025} had newly started to use the Gemini app, and people had created \href{https://x.com/joshwoodward/status/1963627742618165270?s=19}{more than 200 million} images. That figure has grown to more than 50 billion images generated as of \href{https://x.com/NewsFromGoogle/status/2056784211181461614}{May 2026}\nocite{newsfromgoogle2026}.

Along with greater reported use, journalists, researchers, celebrities, and regular people have shared new and/or more detailed \emph{ways} to use image generation tools: from celebrities and influencers sharing LLM-generated images of them \href{https://www.yahoo.com/entertainment/celebrity/articles/kim-kardashian-fans-tears-ai-102147629.html}{hugging their younger selves}\nocite{yahooent2025}; to journalist suggestions that diagrams and infographic capabilities could become a part of \href{https://www.rdworldonline.com/googles-new-nano-banana-pro-could-push-image-generation-deeper-into-rd-and-engineering-workflows/}{research \& development and engineering workflows}\nocite{rdworld2025}; to veterinary science educators creating breed-specific imaging (e.g., brachycephalic airway anomalies in Bulldogs) in an effort to reduce cadaver dissections \citep{choudhary2026}.

Beyond proactive self-reports in traditional media and on social networks, there is limited data on \emph{how} and \emph{why} many people are using LLM image generation. With this growth in usage and emergent use cases, we sought to provide some descriptive analysis of Nano Banana use, to help understand the ways in which users might find tools like Nano Banana helpful for social connections, educational materials, scientific advancement, commerce, and more.

But images of people (and their loved ones)---whether uploaded to Gemini as reference images or generated with Nano Banana---are among the most sensitive data that a person can share. We sought to explore whether we could accurately describe why, how, and how often people are using Nano Banana while protecting the privacy of individual users---to summarize use without ever seeing a single reference image or generated output, and without looking at individual prompts with human eyes. We sought to report on Nano Banana use, with formal end-to-end differential privacy (DP) guarantees; all aggregate statistics together use a total privacy budget of $\varepsilon \le 0.04$ (Section~\ref{sec:privacy_accounting}).

Recently, \citet{liu2025urania} reported on the \href{https://research.google/blog/a-differentially-private-framework-for-gaining-insights-into-ai-chatbot-use/}{Urania} framework and tool for generating insights about LLM chatbot interactions with rigorous DP guarantees. Differential privacy \citep{dwork2006} is considered the gold standard for privacy protection in data analytics and machine learning \citep[see, e.g.,][]{desfontaines2021}. With a private clustering mechanism and multiple keyword extraction methods, benchmarking evaluations showed that Urania could extract meaningful insights about chatbot conversations while maintaining user privacy. Here, we seek to apply these methods to Gemini conversations where users utilize image generation and editing tools: with privacy preserving summarization, clustering, and keyword extraction, can we accurately describe what people are asking LLMs to do with images, and how?

\section{Background and related work}
\label{sec:related_work}

Our work builds on the history of R\&D in image generation models \citep[see][]{shirvani2026} along with research in multiple disciplines which have attempted to improve image generation and evaluation for a variety of purposes, and identify use cases for LLM image generation and editing in general and in specific domains.

\noindent\textbf{Privacy Preservation and Evaluation of LLM Use.} Topic modeling is a major field of studies for over thirty years. However, until recently privacy-preserving methods for this field were not well-developed. The need for summarizing and visualizing insights from user conversations with LLM chatbots creates the incentive to develop such methods alongside LLMs themselves. The recent publication of CLIO \citep{tamkin2024}, a system that uses LLMs to aggregate insights while preserving user privacy, building on \citet{zheng2023}; \citet{zhao2024}, provides aggregate insights about how \href{https://claude.ai}{Claude.ai} is used in the real world, and visualizes aggregate insights. \citet{handa2025} recently applied CLIO to provide insight into the types of economic tasks (classified by O*NET) performed in ways augmented or replaced by chatbot help. More recently, \citet{iscenko2026} developed an ATLAS of 15 million de-identified interactions from LLMs to make observations about diffusion of LLM tools and use in day-to-day and work life. Finally, \citet{liu2025urania}, \citet{cohen2026}, and \citet{chen2026ship} proposed differentially private algorithms for the task.

\noindent\textbf{Analyzing Images at Scale and with LLMs.} Researchers have looked at sources of noise and bias in image training sets, and ways to minimize the limitations of LLM image training data and models \citep{crawford2021}. There is a large literature on cryptography and protection of privacy for images \citep[see, e.g.,][]{wen2024}. \citet{sinha2024} use LLM-powered analysis to look at e-commerce images and attempt to identify fashion trends and attributes. To take a specific domain for a more detailed example, several researchers have examined ways to conduct analysis of medical images (e.g., x-rays), whether through model development, use of public and anonymized training data sets, and privacy-preserving analytic techniques \citep{zhu2025, kaissis2020}. Different from our analysis, these usually involve analyzing the images themselves, or images and text, and for a single diagnosis, as opposed to sample or population estimates.

\noindent\textbf{LLM Image Generation in Education.} Many researchers have innovated and documented approaches to incorporating LLM image generation in educational settings. \citet{feng2026} provide a co-creation system for educational infographics. \citet{kibar2024} describes infographic generation as essential for education of Gen Alpha students. \citet{zhuang2025} use LLMs in language learning to provide learners with LLM-generated real life visual description tasks to prompt learners to describe what they see (and provide multi-modal feedback). \citet{smith2025} describe a way to prompt 8--12 year old learners to sketch their understandings of scientific concepts, and use LLMs to analyze and assess their sketches. \citet{valverderebaza2024} use LLM generated graphics to support data science and analytics education, finding participants preferred LLM generation (and also finding variation in learning curves). \citet{wen2025} propose using multimodal LLMs to generate reviews and surveys of literature topics, including with images and graphical summaries.

\noindent\textbf{LLM Image Generation in Marketing \& Branding.} The application of LLM image generation for marketing and branding purposes has been the subject of research and speculation. \citet{liu2025marketing} evaluate a system to use LLMs to generate---and evaluate before use---marketing and branding materials for online advertising. \citet{chen2025marketing} note the multimodal use case needs for digital marketing and evaluate LLM options. \citet{arora2025} posit LLM-human hybrid models for creating market research plans. Other researchers describe ways to use LLMs to generate response scripts in telemarketing conversations \citep{wang2024} or to generate brand collateral and logos \citep{gawad2025}.

\noindent\textbf{LLM Image Generation in E-Commerce.} \citet{nulli2026} find customization of vision LLMs for e-commerce can result in stronger insights and better outcomes. \citet{iyer2024} describe case studies of uses of LLMs in e-commerce, including visual search and generation of assets. \citet{vashishtha2024} describe LLM-based methods to produce personalized e-commerce banners. \citet{yang2025} describe an approach balancing creativity and conversion goals in prompting to produce stronger e-commerce results from LLM-generated materials. Xingye Chen et al.~\citeyearpar{xingyechen2025} used a reward system to optimize LLM image generation for e-commerce click through rates. Not all documented uses of LLMs for e-commerce have been uniformly positive: Ziwei Chen et al.~\citeyearpar{ziweichen2026} analyzed roughly 1,300 LLM-generated web components and found that more than half contained at least one deceptive design element.

\noindent\textbf{Surveys on LLM Usage.} Another way to approach the question of how people are using AI is simply to ask them, something that public opinion firms and researchers have been doing \citep[see, e.g.,][]{pew2026}. These data and reports are useful, but they are frequently high level and relatively low sample size, and are limited by restricted survey length, recall bias in self-reports, and the increasing reports of AI chatbots and agents in survey panels \citep[see, e.g.,][]{hitches2025}.

We combine aspects of these approaches to answer two narrow questions: (1) Can we, in a privacy preserving way, identify patterns in the use of LLM image generation and editing tools for social, educational, marketing, commercial, and spatial design uses, and (2) identify trends and differences in \emph{how} people are using the tools for those purposes?

\section{Methods, sample, \& analysis}
\label{sec:methods}

Our challenge in answering these questions is that we would not actually be able to view individual images and conversations for this analysis. Images---sometimes of users' families or themselves, sometimes of users' creations or products---are among the most private things users upload, or talk about. Google is a cautious steward of user data, and data access and use restrictions are especially high for users' images. We could not do this analysis if it required direct analysis of Nano Banana images themselves. However, we have been developing tools to balance privacy of users' data, with a high degree of quality of aggregate insights. By applying the \href{https://research.google/blog/a-differentially-private-framework-for-gaining-insights-into-ai-chatbot-use/}{Urania framework} to conversations where users ask for Nano Banana image generation or editing, we can, with end-to-end privacy protection, understand aggregate trends in what people are asking Nano Banana to do.

\subsection{Urania---a privacy preserving method of analyzing text LLM prompts}
\label{sec:urania}

At a high level, Urania uses a \href{https://research.google/blog/practical-differentially-private-clustering/}{private clustering algorithm} on the conversations and releases summaries of the conversation clusters. In addition to clustering, we ask multiple choice questions about each conversation and aggregate them within the conversation categories we define. Importantly, both the clustering algorithm and the follow-up questions use the gold standard for private data analysis, called differential privacy (DP). This ensures that, in a mathematically provable sense, the output summaries and multiple choice aggregates do not reveal information about any individual conversation.

\begin{figure*}[htbp]
  \centering
  \includegraphics[width=0.78\textwidth]{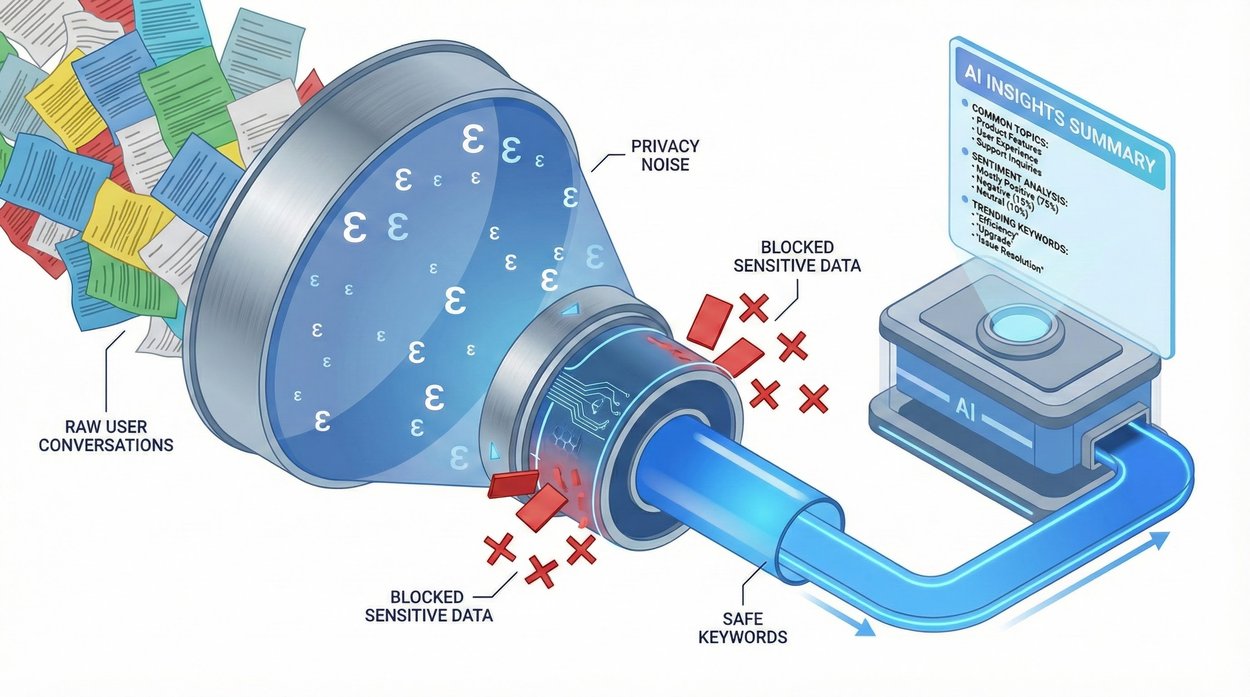}
  \caption{Visualization (generated by Nano Banana) of Urania framework for summarizing chatbot conversations.}
  \label{fig:urania_viz}
\end{figure*}

\begin{figure*}[htbp]
  \centering
  \includegraphics[width=0.70\textwidth]{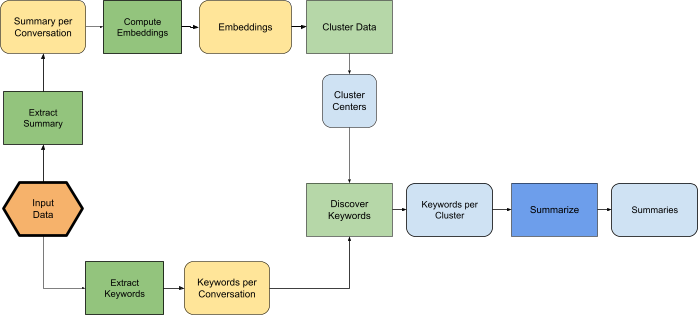}
  \caption{The framework's data flow. Yellow nodes denote non-DP data, green nodes represent operations that are either DP or per conversation, light blue nodes denote private data, and dark blue nodes represent non-private operations.}
  \label{fig:urania_dataflow}
\end{figure*}

By integrating DP at its core, this framework's privacy guarantees are mathematical, not heuristic. They don't depend on an LLM's ability to perfectly redact private data: in other words, even if keywords contain PII or some other sensitive data, the generated summaries are not going to contain this data. In more practical terms, this guarantee makes it impossible for the LLM to reveal sensitive data during summarization and analysis (e.g., due to prompt injection attacks).

Urania's evaluations and comparative benchmarks \citep{liu2025urania} show both that the Urania method protects privacy (a membership inference-style attack showed minimal data leakage compared to alternatives), and has benchmark-leading ability to extract meaningful conversational insights (evaluators preferred Urania-generated cluster summaries from general frequent keywords to non-private summaries).

\subsection{Sample}
\label{sec:sample}

For the purposes of  privacy preservation, we selected a random sample of Gemini conversations, worldwide, from 90 days in Q1 2026, which used Nano Banana for one or more prompts. We used Urania to develop a suggested taxonomy of intents and uses, and filtered to the following five high-level categories. A sample of 2.1 million conversations of interest, with over 5.2 million individual prompts, was classified at the conversation-level into one of five high level categories of content and intent:

\begin{enumerate}
  \item \textbf{Brand \& Marketing Collateral}: ``Intent is professional brand development and external business communication. Prompts typically involve commercial growth, public-facing identity, or marketing layouts that require brand consistency and legible messaging.''
  \item \textbf{Educational \& Technical Information Media}: ``Intent is pedagogical, procedural, or informational. Focuses on simplifying complex data or processes into clear visual formats for the purpose of teaching, training, or global accessibility.''
  \item \textbf{Product Visualization \& E-Commerce}: ``Intent is the realistic and professional showcase of physical goods. Focuses on the physicality of objects for sales, prototyping, or e-commerce display, requiring specific lighting, materials, and staging.''
  \item \textbf{Architectural \& Spatial Design}: ``Intent is the technical visualization of built environments and layouts. Prompts involve structural logic, interior/exterior renders, and environmental planning meant for professional or industrial use.''
  \item \textbf{Recreation \& Personal Expression}: ``Intent is hedonic, leisure-based, or personal creative exploration. Prompts are for entertainment, social media sharing, or private hobbies without intended professional utility or commercial output.''
\end{enumerate}

We sought to compare cases where the use cases were as clear as possible. Because of this, we remove prompts and conversations where there is not enough information to determine a categorization, or where they were classified as ``none of the above.'' We then further manually reviewed DP-sanitized cluster titles and summaries in the Recreation \& Personal Expression category, and filtered to only those clusters which matched with social media trends we identified from \href{https://www.fox10tv.com/2026/02/04/ai-caricatures-latest-online-trend/}{traditional} \href{https://timesofindia.indiatimes.com/etimes/trending/ghibli-and-nano-banana-the-crazy-ai-trends-that-took-over-the-internet/articleshow/123842688.cms}{media}, \href{https://www.techradar.com/ai-platforms-assistants/gemini/i-turned-myself-into-a-3d-figurine-with-googles-nano-banana-heres-how-you-can-hop-on-the-latest-ai-image-trend}{tech press}, \href{https://www.reddit.com/r/GeminiAI/comments/1q2puu2/i_tested_100_nano_banana_pro_prompts_and_narrowed/}{social} \href{https://www.instagram.com/p/DOqdV8sk7_0/?hl=en}{media}, and \href{https://www.canva.com/ai-art-generator/global-trends/}{external} and \href{https://blog.google/products-and-platforms/products/gemini/nano-banana-google-trends-2025/}{internal} blog posts. Within the sample of 5.2 M prompts, we selected the following groups of clusters from the Urania analysis. We then excluded clusters $<50$ conversations for additional privacy protection. This left us with sets of clusters within these five topics:

\begin{itemize}
  \item `Viral / Social' clusters: 375
  \item `Brand \& Marketing Collateral' clusters: 191
  \item `Product Visualization \& E-Commerce' clusters: 135
  \item `Architectural \& Spatial Design' clusters: 66
  \item `Educational \& Technical Information Media' clusters: 64
\end{itemize}

Aggregate information on the default languages and home countries reported by users showed concentrations of use of Nano Banana in Q1 2026 in some larger countries (e.g., US, India, Brazil) and more common languages (e.g., English, Spanish, Arabic, Portuguese)---but also a wide variety of languages and home countries in the sample.

\begin{figure*}[tbp]
  \centering
  \includegraphics[width=0.82\textwidth]{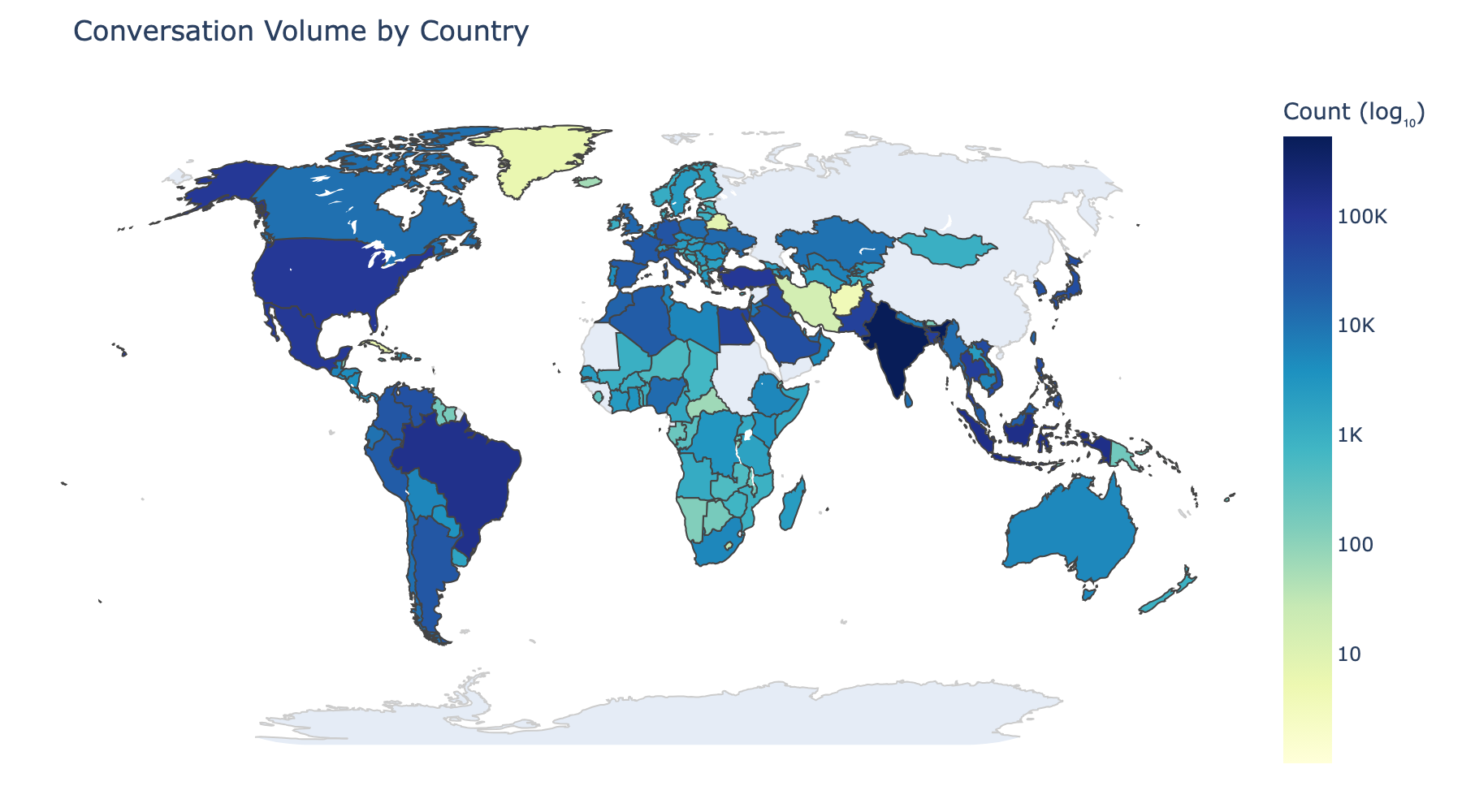}
  \caption{Conversation volume distribution by country.}
  \label{fig:volume_country}
\end{figure*}

\begin{figure*}[tbp]
  \centering
  \includegraphics[width=0.82\textwidth]{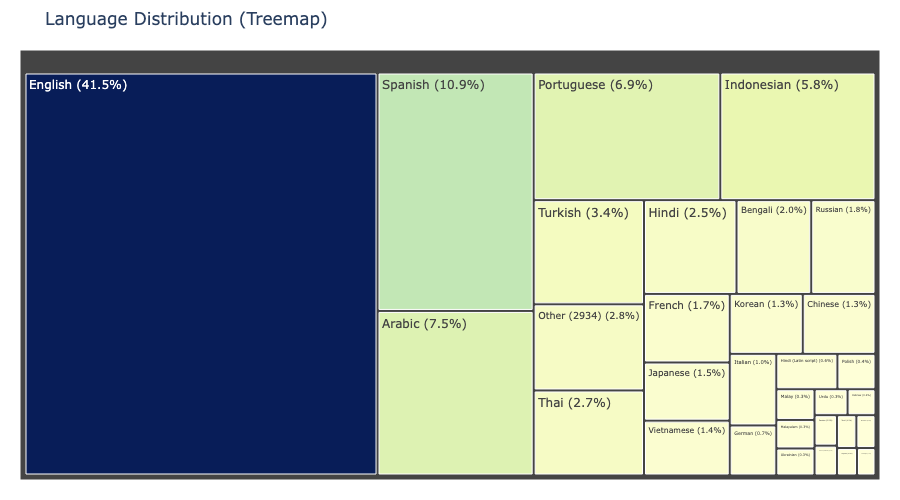}
  \caption{Conversation volume distribution by language.}
  \label{fig:volume_language}
\end{figure*}

\subsection{Clustering within High-Level Categories, \& Assigning Feature Ratings}
\label{sec:clustering}

We used Urania to group prompts into clusters underneath the five high level categories. The clusters were identified through mathematical embeddings and summarized by Urania.

We additionally used Urania and the Gemini API to rate conversations on 21 binary \emph{features}, giving each conversation a yes/no value on various ``how'' questions in four categories. \emph{Image Editing} included explicit requests for modifications to alter, clean up, or change specific elements within an image that has already been created (or a reference image uploaded by a user). \emph{Artistic Style} included explicit user requests about the creative direction and visual aesthetic, from the artistic medium used to specific stylistic or cultural influences. \emph{Composition} refers to requesting specifics about the structural arrangement of the scene, including placement of subjects, perspective, lighting, etc. \emph{Interactions} features look at the conversation as a whole and how people iterate on and edit generated images.

\subsubsection*{Image Editing Features}
\begin{itemize}
  \item \textbf{Remove Object}: The user explicitly requests removing specific objects, blemishes, or artifacts from the image.
  \item \textbf{Remove People from Image}: A request to erase human subjects from the scene.
  \item \textbf{Modify Background}: User explicitly requests changes or edits to the background of the image, or removing the background, or specific elements in a background.
  \item \textbf{Extend, Expand, or Fill}: User requests adding new elements, expanding the canvas (outpainting), or replacing a masked area with new generated content.
  \item \textbf{Enhance or Resize}: User requests changing the aspect ratio, resolution, or enhancing the overall crispness/size of an existing image.
  \item \textbf{Restore or Repair}: The user requests repairing old, damaged, blurry, or corrupted images to their original or improved quality.
\end{itemize}

\subsubsection*{Artistic Style Features}
\begin{itemize}
  \item \textbf{Artistic Medium}: Requests specifying a particular art material or format (e.g., watercolor, oil painting, pencil sketch, charcoal, digital art, photography).
  \item \textbf{Vibe or Style}: Requests defining the aesthetic mood, genre, or specific stylistic direction (e.g., ``cyberpunk,'' ``minimalist,'' ``vintage,'' ``surrealist,'' ``film noir'').
  \item \textbf{Conceptual Mashup}: Combining two or more disparate concepts, objects, or themes to create a novel image.
  \item \textbf{Absurdity or Humor}: Requests that intentionally incorporate elements of nonsense, satire, surreal comedy, or humorous exaggeration.
  \item \textbf{Imitate Art, Person, or Character}: Explicit requests to emulate the distinct style of a specific artist, public figure, or fictional character.
  \item \textbf{Meme Style or Format}: Creating content that adopts the visual structure, layout, or cultural references typical of internet memes (e.g., ``Drake Hotline Bling format,'' ``two-panel meme'').
\end{itemize}

\subsubsection*{Composition Features}
We looked at the following features of use across our five use cases of interest:
\begin{itemize}
  \item \textbf{Modify Lighting}: Adjusting the light source, intensity, color temperature, time of day, or creating effects like ``dramatic shadows'' or ``soft, diffused lighting.''
  \item \textbf{Perspective or Angle}: Changing the viewpoint of the scene (e.g., ``bird's eye view,'' ``worm's eye view,'' ``close-up,'' ``wide angle'').
  \item \textbf{Precise Positioning}: Specifying the exact location or layout of elements within the frame (e.g., ``put the dog on the left,'' ``centered composition'').
  \item \textbf{Include Multiple Subjects}: Ensuring multiple distinct characters or objects are present within the generated image.
  \item \textbf{Negative Constraints}: Explicit instructions to exclude specific objects, colors, styles, or concepts from the output.
  \item \textbf{Includes People}: Request references a person or people in the image, or asks for person or people to be added.
\end{itemize}

\subsubsection*{Interactions / Session Features}
\begin{itemize}
  \item \textbf{Full Reprompt}: Discarding the previous output and generating a new image from scratch using the same or nearly identical prompt.
  \item \textbf{Simple Iteration}: Asking for small or simple changes to a previous round of image generation or editing (e.g., ``make it slightly darker,'' ``do it again'').
\end{itemize}

As part of privacy preservation, these Yes / No ratings are reported at the cluster level. Each cluster has a mean percentage of conversations that included, e.g., Removing Objects, or asking to Modify or specify lighting.

This allows us to simply compare the intents and contents---the types of clusters summarized in each category---and also the ways people asked for content, or how people used the tool for different purposes.

\subsubsection*{Multi-select Features}
\label{sec:multi_select}

In addition to these binary feature ratings, we also used Urania to rate on a set of multi-select features, including:

\paragraph{Text and Typography.}
Which best describes the user's intent regarding specific text, lettering, or typography within the image?
\begin{itemize}
  \item \emph{Full Specification}: The user provides exact text strings (to add, change, or remove) AND specifies a visual style, font, or effect (e.g., `the word SALE in neon script', `change John to Mike in bold red letters').
  \item \emph{Precise Content Only}: The user provides exact words or characters to be rendered, changed, or deleted, but does NOT specify a visual style (e.g., `write the name John', `change the word to OFF', `remove the text from the sign').
  \item \emph{Stylistic/Generic Only}: The user describes the `look', font, or `vibe' of lettering WITHOUT providing specific words (e.g., `graffiti style', `serif font', `bold 3D typography').
  \item \emph{N/A}: The prompt contains no mention of text, letters, or typography.
\end{itemize}

\paragraph{Asset Transformation \& Utility Type.}
Does the user's request focus on modifying, cleaning, or auditing an existing visual asset/concept rather than generating a whole new scene from scratch?
\begin{itemize}
  \item \emph{Content Removal/Cleanup}: Removing specific elements like watermarks, background clutter, logos, or `AI artifacts' to clean up the image.
  \item \emph{Modification/Transformation}: Intentionally changing specific properties of the image, such as swapping a logo, changing the text content, altering the color of specific objects (e.g., `make the shirt blue'), or replacing an item.
  \item \emph{Logical/Technical Correction}: Fixing errors in factual content, such as correcting a chemical structure, fixing spelling in a diagram, or auditing technical logic.
  \item \emph{Style/Quality Enhancement}: General improvements to resolution, color grading, or `pro' aesthetic without changing the specific subjects or text.
  \item \emph{None/New Generation}: A request for a brand-new generation from scratch with no transformative or corrective intent.
\end{itemize}

\paragraph{Follow-up questions rating.}
Individual categories also were rated on \emph{follow up questions} by Urania, such that conversations that were rated into individual clusters were also categorized by features of use specific to those clusters (e.g., educational prompts were rated on follow-up questions about different types of educational use; marketing prompts were rated on follow-up questions about brand identity, etc.).

\subsection{Privacy accounting}
\label{sec:privacy_accounting}

All statistics reported in this paper are released under differential privacy, with a single conversation as the unit of privacy. Conversations are included via Poisson subsampling. The total privacy budget is $\varepsilon \le 0.04$ ($\delta = 0$).

\subsection{Analysis: Comparing Categories by Prompt Linguistic Properties: Word, Noun, and Adjective Counts}
\label{sec:analysis_linguistic}

\begin{figure*}[tbp]
  \centering
  \includegraphics[width=0.98\textwidth]{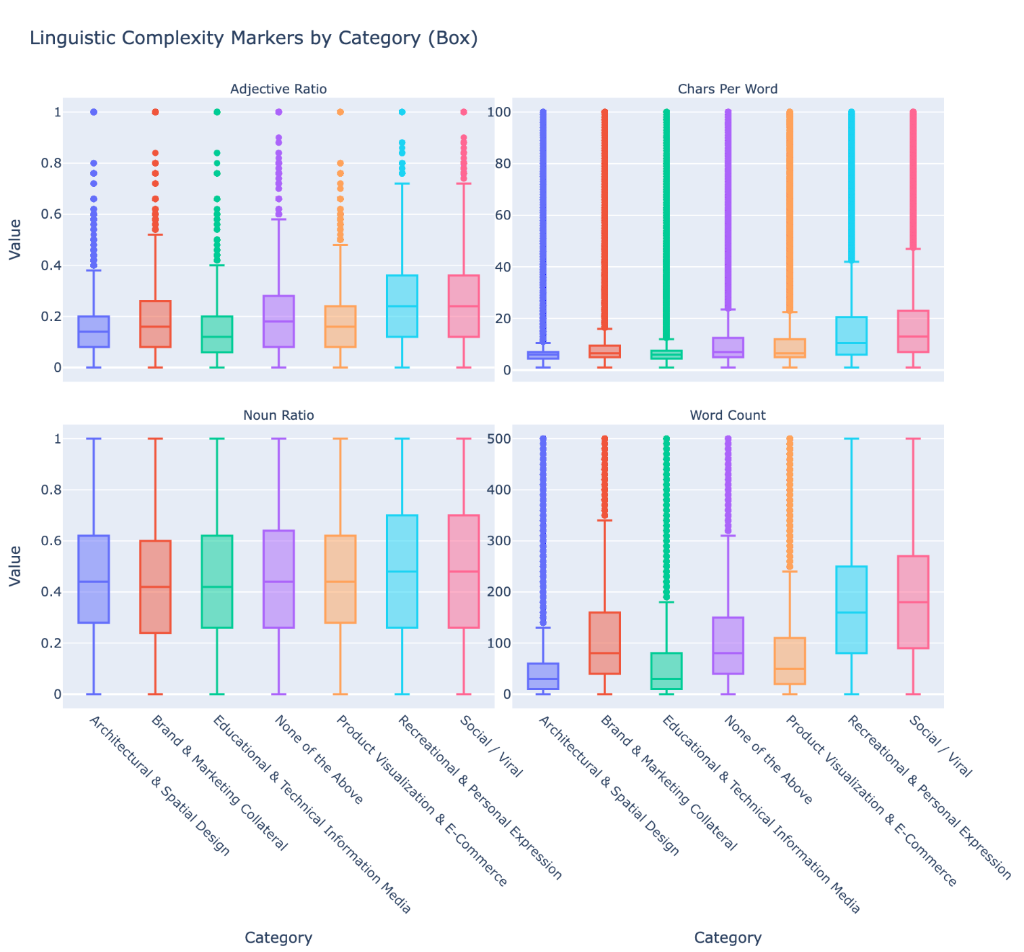}
  \caption{Conversation linguistic properties distribution by category.}
  \label{fig:linguistic_props}
\end{figure*}

In addition to rating on intents and features of use, we also conducted analysis on the linguistic properties of conversations in the different categories of use. In addition to character and word counts, we used the \href{https://spacy.io/}{spaCy Python library} of language definitions to count nouns and adjectives within each conversation. We use differential privacy methods to report in Figure~\ref{fig:linguistic_props} on the distribution of word count, adjectives per total words, nouns per total words, and characters per word for the different high level categories.

The differential privacy protections definitionally introduce more noise into these measures, making it impossible to make precise comparative statements. However, we do see some suggestions of potential differences in the distribution of linguistic complexity across these categories.

\subsection{Analysis: Social Media \& Viral trends in Nano Banana use}
\label{sec:analysis_social}

Social media uses of Nano Banana and other LLM trends have been some of the most frequently reported and shared use cases (e.g., the article ``\href{https://tech.yahoo.com/ai/articles/nano-banana-just-broke-internet-130036798.html}{Nano Banana just broke the internet with these viral trends}''\nocite{yahootech2026}).

Table~\ref{tab:social_clusters} below shows a selection of cluster titles analyzed in the Social \& Viral Image generation clusters. We do see some evidence of social media uses and viral trends referenced in social media and articles appearing in the Nano Banana log summaries.

\begin{table}[htbp]
  \centering
  \caption{Example Social \& Viral Image generation cluster titles.}
  \label{tab:social_clusters}
  \footnotesize
  \begin{tabularx}{\columnwidth}{@{}>{\raggedright\arraybackslash}X@{}}
    \toprule
    \textbf{Example Social \& Viral Image Generation Clusters} \\
    \midrule
    \vspace{-0.6\baselineskip}
    \begin{itemize}[leftmargin=1.1em, itemsep=2pt, topsep=0pt, partopsep=0pt, parsep=0pt]
      \item 19th-Century Botanical Atlas Illustrations with Personal Annotations
      \item Crying with Celebrities at Night in Convenience Stores
      \item Generating images of people in Brazil jerseys
      \item Generating Met Gala Outfits and Red Carpet Looks
      \item Generating Football Broadcast Screenshots with Specific Fans
      \item Transforming Photos into Distinct Cartoon Styles
      \item Polaroid-Style Photos with White Curtain Background
      \item Cinematic Modification: Adding Demonic Smoke Entities
      \item Playful Baldness Transformations in Image Editing
      \item Visualizing Age Progression and Regression on People
      \item Personalized Sports Jersey Edits for Photos
      \item Whimsical Cat Image Generation and Editing
      \item Personalized Birthday and Greeting Cards with Text
      \item Transforming Photos into Artistic Oil Paintings
    \end{itemize}
    \vspace{-0.5\baselineskip} \\
    \bottomrule
  \end{tabularx}
\end{table}

To further examine aspects of what and how people were using Nano Banana, based on those clusters we sought to be able to describe prevalence, by identifying some keywords that we could detect in Urania clusters. For example, you can see in Table~\ref{tab:social_clusters} the cluster ``Transforming Photos into Artistic Oil Paintings.'' We identified a list of the 200 most frequently searched artists in Google Trends data, and identified the frequency with which those artist names showed up in Nano Banana prompts. To be clear, nowhere in this analysis do we look at any individual images, and, as a hypothetical example, attempt to rate the images on the presence of surrealist elements. We only count, in a privacy preserving way, the frequency with which users mention Dali in prompts. Table~\ref{tab:artist_keywords} shows the rank order of that frequency.

\begin{table}[htbp]
  \centering
  \caption{Most frequently referenced Artist keywords in Nano Banana prompts.}
  \label{tab:artist_keywords}
  \footnotesize
  \begin{tabular}{@{}r l@{}}
    \toprule
    \textbf{Rank} & \textbf{Artist} \\
    \midrule
     1 & Rembrandt \\
     2 & Salvador Dali \\
     3 & Francisco Goya \\
     4 & Vincent Van Gogh \\
     5 & Leonardo da Vinci \\
     6 & Caravaggio \\
     7 & Pablo Picasso \\
     8 & Edgar Degas \\
     9 & Claude Monet \\
    10 & Edward Hopper \\
    11 & J.M.W.\ Turner \\
    12 & Michelangelo \\
    13 & Andy Warhol \\
    14 & Henri Matisse \\
    15 & Peter Paul Rubens \\
    16 & Banksy \\
    17 & Edvard Munch \\
    18 & Frida Kahlo \\
    19 & Eduardo Kobra \\
    20 & Gustav Klimt \\
    21 & Wassily Kandinsky \\
    22 & Raphael \\
    23 & Johannes Vermeer \\
    24 & Roy Lichtenstein \\
    25 & Jean-Michel Basquiat \\
    \bottomrule
  \end{tabular}
\end{table}

\begin{figure}[htbp]
  \centering
  \includegraphics[width=0.80\linewidth]{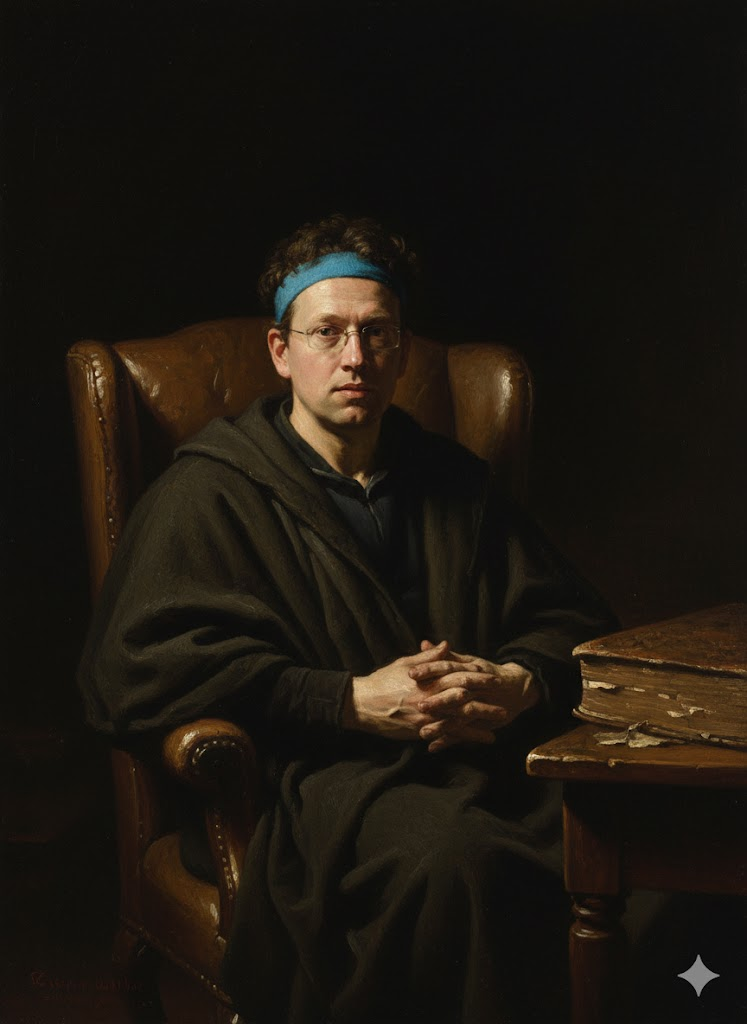}
  \caption{A co-author in Rembrandt-style portrait (included with permission).}
  \label{fig:rembrandt}
\end{figure}

To test them out ourselves, the authors identified public posts where people shared their prompts, and that prompt included, e.g.\ \href{https://121clicks.com/inspirations/gemini-ai-prompts-rembrandt-style-look/}{Rembrandt}\nocite{clicks2026}. We then uploaded a reference image to Gemini along with the prompt, to explore what such a prompt would look like. Figure~\ref{fig:rembrandt} shows the result of that prompt---an example of what some of those social prompts may have looked like---again, we do all this without looking at any individual prompts, and in a privacy preserving manner (in this case, the image is shared with the co-author's permission).

\begin{figure}[htbp]
  \centering
  \includegraphics[width=\linewidth]{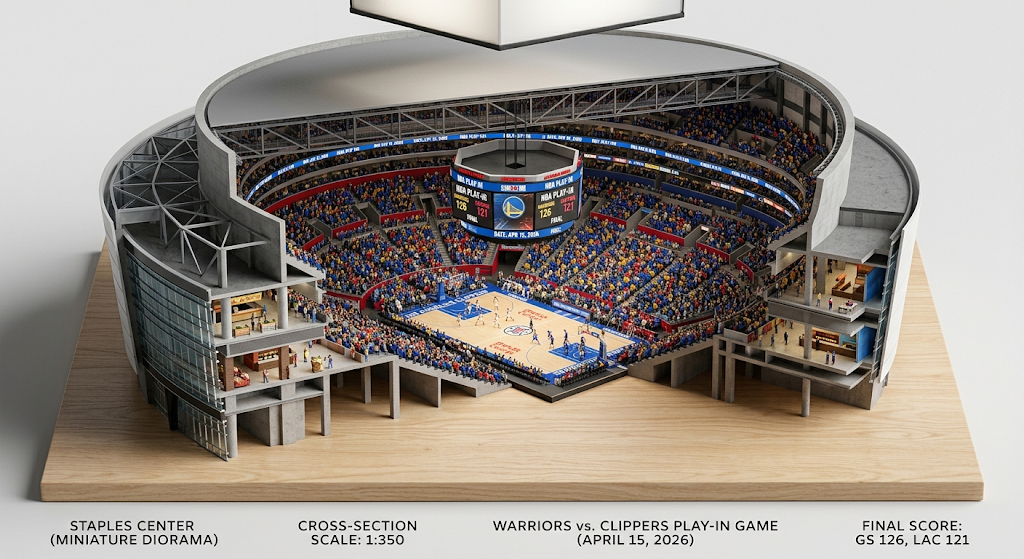}
  \caption[Illustrative social media image based on sports tweet.]{Illustrative social media image based on a sports \href{https://x.com/sundarpichai/status/1996783899943620967?lang=en}{tweet}.}
  \vspace{0.3em}
  {\footnotesize\raggedright\emph{Note:} Based on a December 2025 \href{https://x.com/sundarpichai/status/1996783899943620967?lang=en}{example from social media}, the exact prompt (entered on April 17, 2026) was: ``Research earlier this week's score from the Golden State Warriors v.\ LA Clippers play-in game, then create a 45$^\circ$ top-down photorealistic isometric miniature basketball arena diorama of the Staples Center (with a cross section visible), use soft, refined textures with realistic PBR materials and gentle, lifelike lighting and shadows.'' Other example social media prompts and frequency lists are included in the appendix.\par}
  \label{fig:cricket}
\end{figure}

We also looked into which sports were most frequently referenced in keywords in clusters. Our sample is global, and was sampled leading up to and during global sporting events. Table~\ref{tab:sports_keywords} shows the rank ordered frequency of those sport keywords, and Figure~\ref{fig:cricket} shows an example based on a prompt shared on \href{https://timesofindia.indiatimes.com/technology/social/ind-vs-sa-3rd-odi-create-ai-generated-miniature-stadiums-with-nano-banana-pro-use-this-prompt-suggested-by-sundar-pichai/articleshow/125795665.cms}{social media}\nocite{toi2026}.

\begin{table}[htbp]
  \centering
  \caption{Most frequently referenced sports keywords in Nano Banana prompts.}
  \label{tab:sports_keywords}
  \footnotesize
  \begin{tabular}{@{}r l@{}}
    \toprule
    \textbf{Rank} & \textbf{Sport} \\
    \midrule
     1 & Motorsports \\
     2 & Football (Ambiguous) \\
     3 & Baseball \\
     4 & Soccer \\
     5 & Cricket \\
     6 & Swimming \\
     7 & Basketball \\
     8 & Esports \\
     9 & Tennis \\
    10 & Golf \\
    11 & Jumping Events \\
    12 & MMA \\
    13 & Boxing \\
    14 & Volleyball \\
    15 & Professional Wrestling \\
    16 & Judo/Taekwondo \\
    17 & Cycling \\
    18 & Rugby \\
    19 & American Football \\
    20 & Athletics \\
    21 & Pickleball \\
    22 & Marathon \\
    23 & Billiards \\
    24 & Table Tennis \\
    25 & Surfing \\
    \bottomrule
  \end{tabular}
\end{table}

\subsubsection*{Prevalence of Use of Features in Social and Viral Clusters}

We observe the prevalence of features for conversations classified into Social and Viral clusters. The highest prevalence (at nearly 80\%) is the inclusion of people. Several compositional (multi-subject scene, precise positioning, modify lighting) and artistic style features (vibe or style, artistic medium) are also frequently present in social and viral clusters.

\begin{figure}[htbp]
  \centering
  \includegraphics[width=\linewidth]{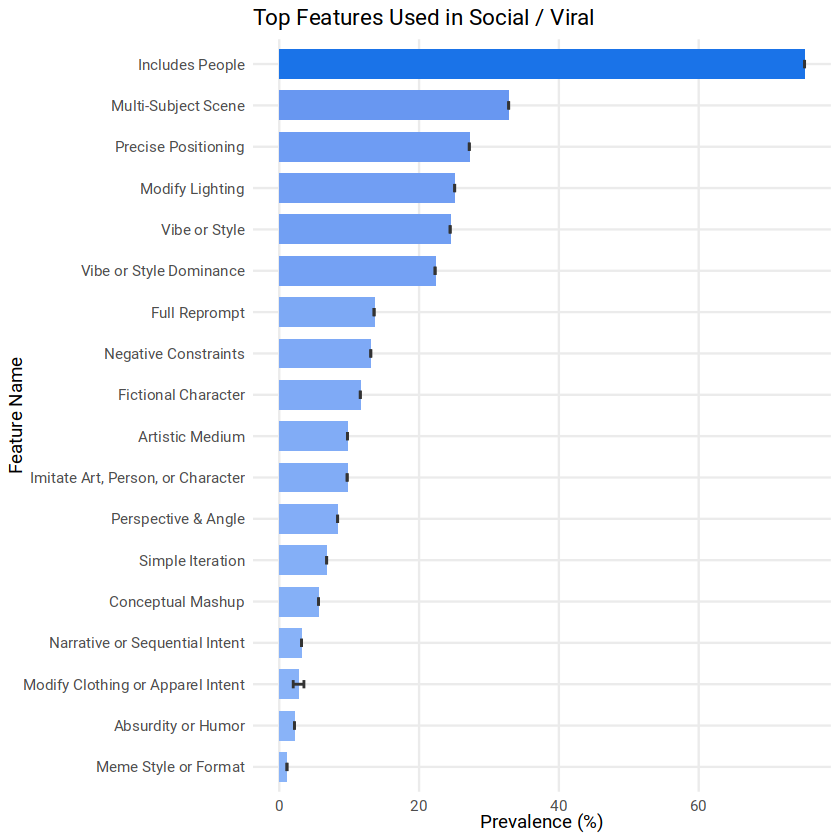}
  \caption{Prevalence of Feature Use in Social and Viral Conversation Clusters.}
  \label{fig:features_social}
\end{figure}

Full reprompts are about twice as prevalent as simple iterations (e.g., ``try again'') in social and viral prompts. Perhaps surprisingly, meme style or format and absurdity / humor are the least prevalent.

\subsection{Analysis: Image Generation and Editing Feature Prevalence in Architectural, Marketing, Educational, and E-Commerce Uses}
\label{sec:analysis_categories}

Here we report on the prevalence of editing and generation features in the clusters of conversations within Architectural, Marketing, Educational, and E-Commerce uses of Nano Banana. In Table~\ref{tab:illustrative_clusters} below, we can see illustrative cluster titles from these high level categories.

\begin{table*}[!t]
  \centering
  \caption[Illustrative Cluster titles from high-level categories.]{Illustrative Cluster titles from categories: Product Visualizations \& e-commerce, Educational \& Technical media, Architectural \& Spatial design, and Brand \& Marketing Collateral.\protect\footnotemark[1]}
  \label{tab:illustrative_clusters}
  \scriptsize
  \begin{tabularx}{\textwidth}{@{}>{\raggedright\arraybackslash}X @{\hspace{1.8em}} >{\raggedright\arraybackslash}X@{}}
    \cmidrule[\heavyrulewidth](r{1.8em}){1-1}\cmidrule[\heavyrulewidth]{2-2}
    \textbf{Product Visualizations \& E-Commerce} & \textbf{Brand \& Marketing Collateral} \\
    \cmidrule(r{1.8em}){1-1}\cmidrule{2-2}
    \vspace{-0.6\baselineskip}
    \begin{itemize}[leftmargin=1.1em, itemsep=1.2pt, topsep=0pt, partopsep=0pt, parsep=0pt]
      \item Cake Photography: Focus on Realistic Digital Staging
      \item Generate Realistic Bag Visuals for E-Commerce
      \item Product Visualization \& Modification of Bottles for E-Commerce
      \item Packaging Design for Product Visualization \& E-commerce
      \item 3D Photorealistic Product Renders with Technical Precision
      \item Product Image Enhancement for E-commerce Display
      \item Product Image Editing: Focus on White Backgrounds
      \item Apparel Visualization on Models and Mannequins
      \item Luxury Jewelry Visualization and Enhancement on Models
      \item Footwear Product Visualization, Modification, and Display
      \item Changing Apparel Colors: Black, White, and More
    \end{itemize}
    \vspace{-0.5\baselineskip}
    &
    \vspace{-0.6\baselineskip}
    \begin{itemize}[leftmargin=1.1em, itemsep=1.2pt, topsep=0pt, partopsep=0pt, parsep=0pt]
      \item Food \& Beverage Logo and Brand Identity
      \item Team Logos \& Branding with Text Modifications
      \item Refining Logos and Visual Identity with Text and Color Edits
      \item Gaming and Esports Logo Design with Characters
      \item Brochure and Pamphlet Design and Creation with Images
      \item Custom Sticker Design and Print Preparation with Text
      \item Custom Business Card Design and Content Edits
      \item Event-Focused Flyer Design and Text Refinement
      \item Restaurant Menu Design with Food Images and Pricing
      \item YouTube Channel Branding: Logos and Banners
      \item Creating World Cup Themed Sports Graphics and Logos
    \end{itemize}
    \vspace{-0.5\baselineskip} \\
    \cmidrule[\heavyrulewidth](r{1.8em}){1-1}\cmidrule[\heavyrulewidth]{2-2}
    \addlinespace[0.85em]
    \cmidrule[\heavyrulewidth](r{1.8em}){1-1}\cmidrule[\heavyrulewidth]{2-2}
    \textbf{Architectural \& Spatial Design} & \textbf{Educational \& Technical Information} \\
    \cmidrule(r{1.8em}){1-1}\cmidrule{2-2}
    \vspace{-0.6\baselineskip}
    \begin{itemize}[leftmargin=1.1em, itemsep=1.2pt, topsep=0pt, partopsep=0pt, parsep=0pt]
      \item Refining Room and Floor Plan Layouts
      \item Precise Technical Drafting for Complex 3D Models
      \item Detailed Generation of Architectural and Technical Spatial Plans
      \item Detailed Floor Plans: Generation, Modification, and Arrangement
      \item Rendering Interior Staircases with Architectural Design Modifications
      \item Visualizing Plant Arrangements and Garden Layouts
      \item Exterior House Color and Modification Iterations
      \item Architectural Rendering of Doors, Gates, and Windows
      \item Architectural Exterior Modification: Structure, Facade, and Elements
      \item Interior Design: Furniture Placement and Room Redesign
      \item Optimizing Interior Layouts and Storage Solutions
    \end{itemize}
    \vspace{-0.5\baselineskip}
    &
    \vspace{-0.6\baselineskip}
    \begin{itemize}[leftmargin=1.1em, itemsep=1.2pt, topsep=0pt, partopsep=0pt, parsep=0pt]
      \item Visualizing and Explaining Historical and Astrological Concepts
      \item Detailed Medical and Scientific Illustration Requests
      \item Visualizing Complex Organic Chemistry Structures and Reactions
      \item Visualizing Mathematical Concepts and Problem-Solving
      \item Physics Diagrams and Scientific Problem Visualization
      \item Generating Educational Materials and Worksheets on Demand
      \item Educational Poster Creation for Class Projects
      \item Step-by-Step Image Editing and AI Generation
      \item Video Editing and AI Creation Guides
      \item Fitness and Diet: Personalized Exercise and Posture Correction
      \item Hands-on Guides for Repairing and Maintaining Physical Items
    \end{itemize}
    \vspace{-0.5\baselineskip} \\
    \cmidrule[\heavyrulewidth](r{1.8em}){1-1}\cmidrule[\heavyrulewidth]{2-2}
  \end{tabularx}
\end{table*}
\afterpage{\footnotetext[1]{As a reminder, these are the definitions the Urania model used to classify the conversations: \textbf{Architectural \& Spatial Design}: ``Intent is the technical visualization of built environments and layouts. Prompts involve structural logic, interior/exterior renders, and environmental planning meant for professional or industrial use.'' \textbf{Brand \& Marketing Collateral}: ``Intent is professional brand development and external business communication. Prompts typically involve commercial growth, public-facing identity, or marketing layouts that require brand consistency and legible messaging.'' \textbf{Educational \& Technical Information Media}: ``Intent is pedagogical, procedural, or informational. Focuses on simplifying complex data or processes into clear visual formats for the purpose of teaching, training, or global accessibility.'' \textbf{Product Visualization \& E-Commerce}: ``Intent is the realistic and professional showcase of physical goods. Focuses on the physicality of objects for sales, prototyping, or e-commerce display, requiring specific lighting, materials, and staging.''}}

\subsubsection*{Architectural and Spatial Design Conversations: Prevalence of Features}

Among the conversations categorized into ``Architectural \& Spatial Design'' prompts, we analyzed the prevalence of binary image editing and generation features. As might have been expected, in these images from clusters such as ``Refining Room and Floor Plan Layouts'' and ``Interior Design: Furniture Placement and Room Redesign'', we see the highest prevalence rates in compositional features like Precise Positioning, Modify Lighting, and Multi-Subject Scenes.

\begin{figure}[htbp]
  \centering
  \includegraphics[width=\linewidth]{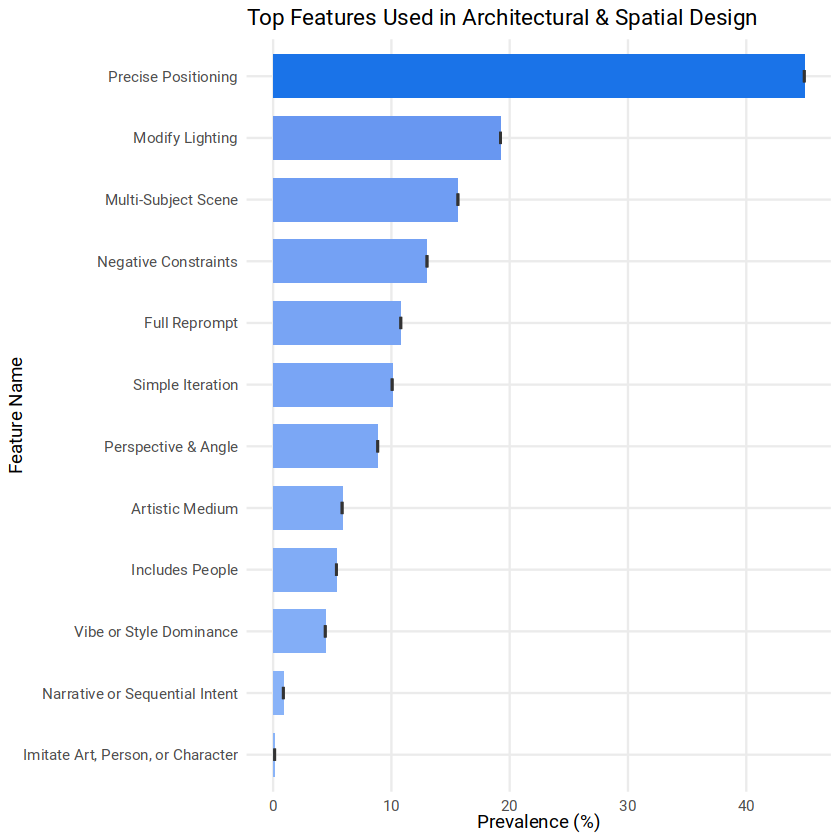}
  \caption{Prevalence of features in Architectural \& Spatial Design conversations.}
  \label{fig:features_arch}
\end{figure}

We also see substantive use of negative constraints, and artistic medium instructions, along with iteration instructions. Meanwhile, architectural and spatial design conversations seem least likely to include fictional characters, vibe or style requests, narrative intent, or people.

\subsubsection*{Brand and Marketing Collateral Conversations: Prevalence of Features}

Among the conversations categorized into Brand and Marketing Collateral clusters, we observed the prevalence of binary image editing and generation features---clusters with titles such as ``Food \& Beverage Logo and Brand Identity,'' ``Gaming and Esports Logo Design,'' and ``YouTube Channel Branding.''

The most prevalent are compositional features: precise positioning, including people, modifying lighting, removing or ensuring absence (negative constraints).

\begin{figure}[htbp]
  \centering
  \includegraphics[width=\linewidth]{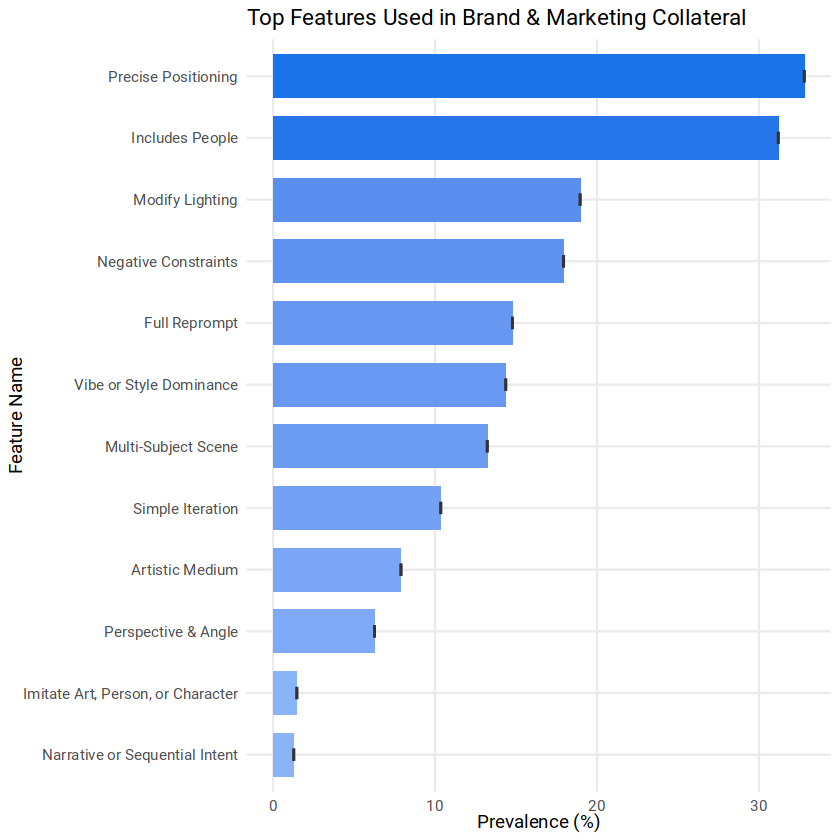}
  \caption{Prevalence of features in Brand and Marketing Collateral conversations.}
  \label{fig:features_brand}
\end{figure}

There are relatively few fictional characters or narrative or sequential intent, and absurdity, humor and memetic structure did not register.

For brand and marketing collateral, we also asked a follow-up question of the sub-categories of clusters, about whether the brand discussed in the conversation was one Gemini recognized as an existing brand or an original or fictional brand.

\begin{figure*}[tbp]
  \centering
  \includegraphics[width=0.58\textwidth]{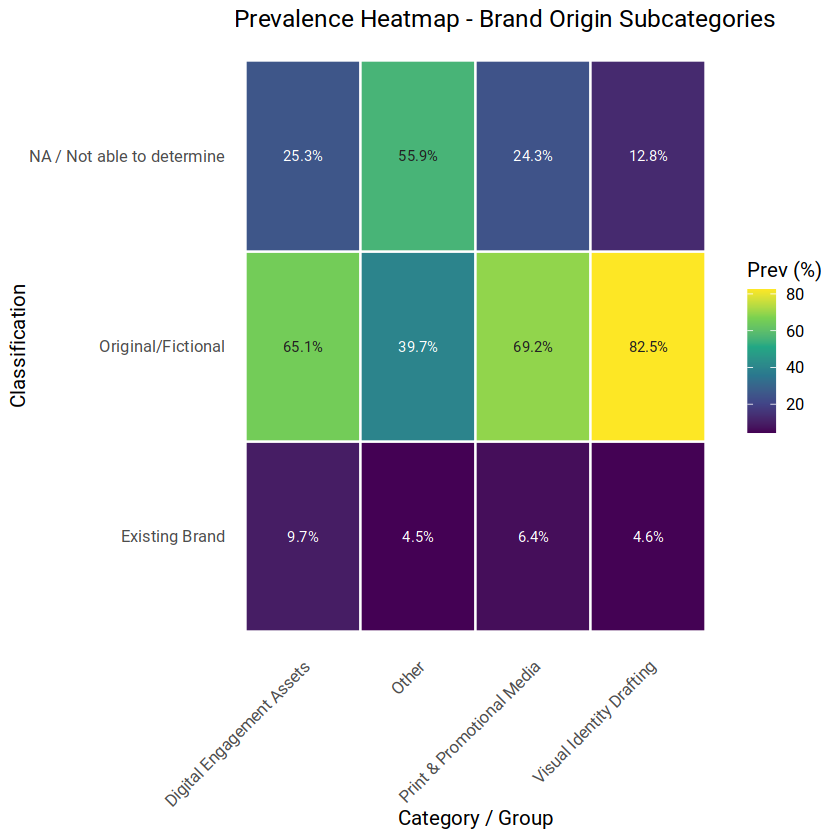}
  \caption{Prevalence of Existing or Original Brands referenced in Brand and Marketing Collateral clusters.}
  \label{fig:existing_brands}
\end{figure*}

A significant proportion of these prompts are rated ``NA,'' but we do see that relatively few seem to be from existing brands recognized by Gemini, and the majority of Digital Engagement Assets, Print \& Promotional Media, and Visual Identity Drafting is for original, fictional, or brands that Gemini does not recognize---suggesting that some users may be developing brand and marketing collateral for new or speculative ventures.

\subsubsection*{Educational \& Technical Information Media Conversations: Prevalence of Features}

We used Urania to classify conversations into Educational \& Technical Information Media clusters, such as ``Physics Diagrams \& Scientific Problem Visualization'' and ``Hands-on Guides for Repairing and Maintaining Physical Items.'' We observed the prevalence of binary image editing and generation features.

Compositional features (Precise Positioning, Multi-Subject Scene, Negative Constraints) are among the most prevalent, though it is perhaps surprising to see ``Includes People'' as the most prevalent feature in Educational \& Technical Information Media. This could be because this category includes clusters of scientific (e.g., anatomy) educational guides, step-by-step troubleshooting or repair guides which may include, e.g., a hand using a piece of equipment, or for other reasons.

\begin{figure}[htbp]
  \centering
  \includegraphics[width=\linewidth]{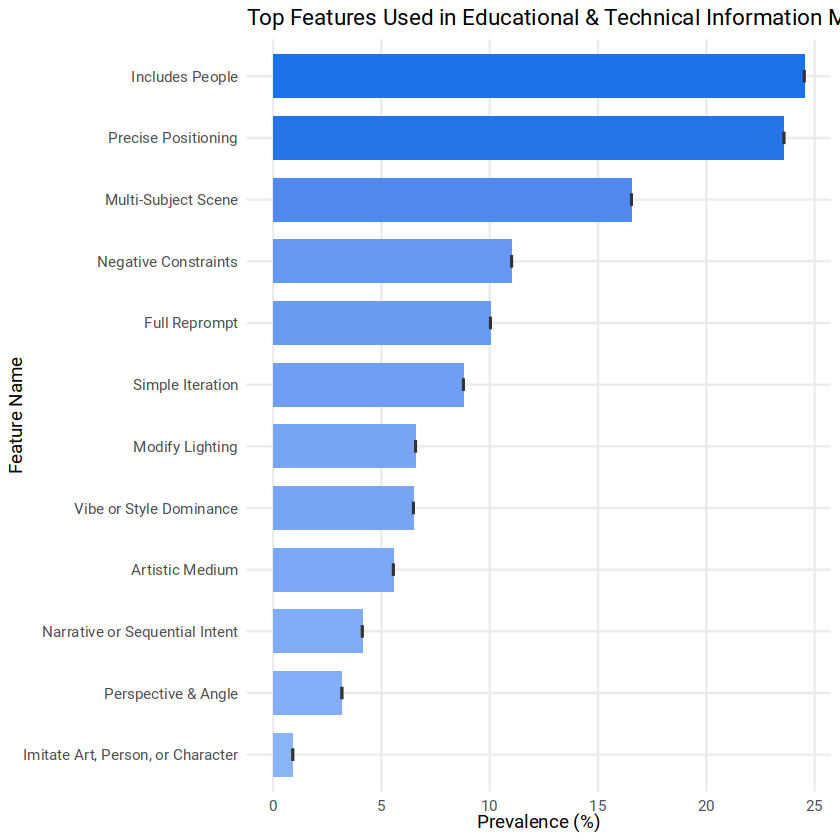}
  \caption{Prevalence of features in Educational and Technical Information Media conversations.}
  \label{fig:features_edu}
\end{figure}

Fictional Character is very low, and Absurdity and Humor and Memetic structure are so low as to not register in this sub sample.

For Educational and Technical graphics, we also asked follow up questions, looking at the educational role of the prompter (if detectable), the domain / subject, the level of abstraction (Diagram/Abstract v.\ Realistic Representation), and the level of the target audience (K--12, Advanced, or Vocational), as possible. The model was instructed to rate as other or NA frequently and only classify into the subgroups when there was high confidence.

As seen in Figure~\ref{fig:edu_roles} below, for the sub-categories of Educational Diagrams \& Infographics and Linguistic and Global Communication Aids, there are a relatively higher proportion of educator and educational content creator roles identified. For Procedural \& Workflow Aids, and Scientific \& Academic Illustrations, the vast majority of roles identified were of students and individual researchers. Many conversations were rated NA, as not having an explicit role identified.

\begin{figure*}[tbp]
  \centering
  \includegraphics[width=0.65\textwidth]{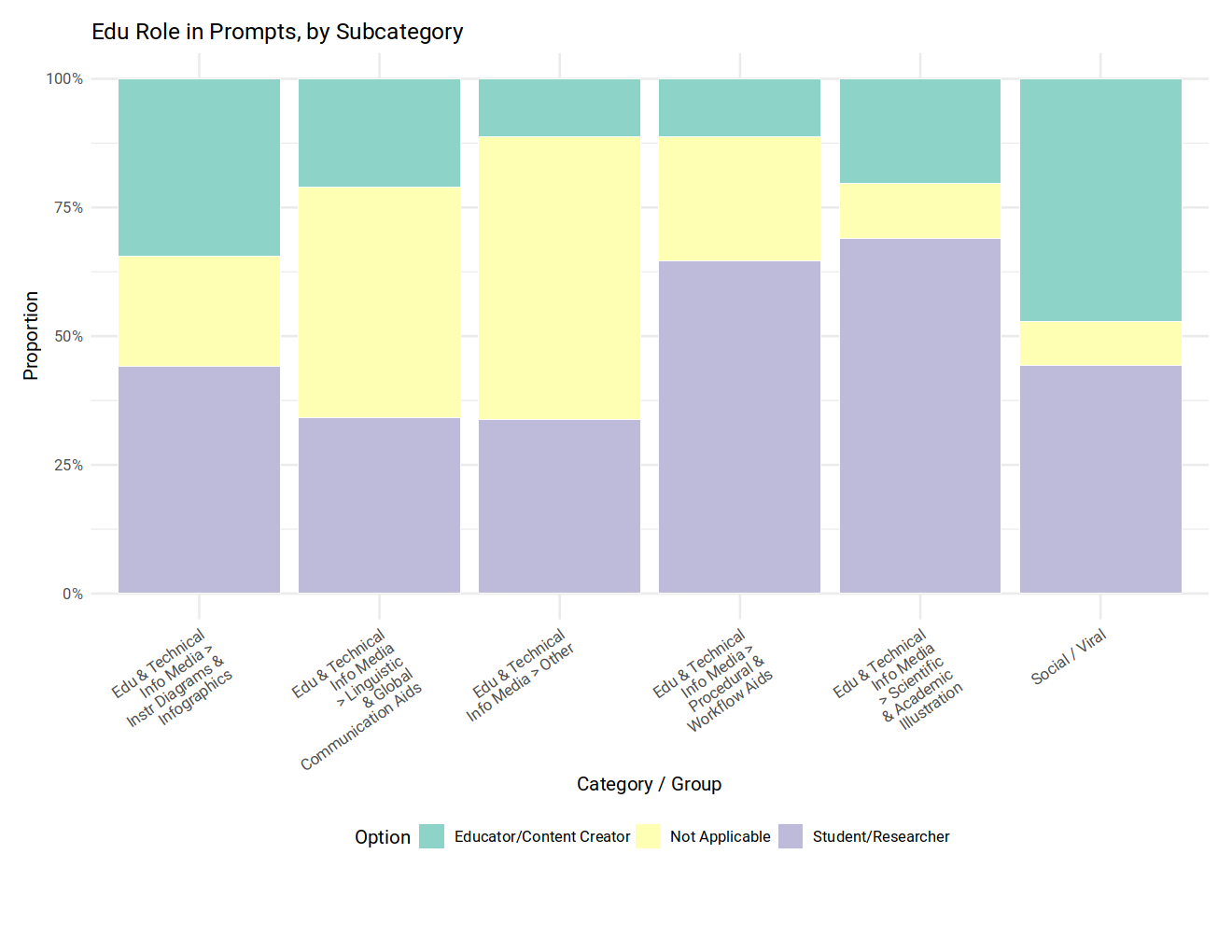}
  \caption{Prevalence of Educator Roles referenced in Educational \& Technical Information Media clusters.}
  \label{fig:edu_roles}
\end{figure*}

Where possible, educational prompts were also classified into subject or domain areas---Global Languages, Social Studies, Science, Vocational/Professional, Math, English / Language Arts, and Art. While a plurality of educational image prompts were not able to be identified by the model as one of these subjects, some trends are visible in the ratings. Linguistic Communication Aids had the highest proportion in the Global languages subject. Academic illustrations are most often in Science subjects. And Procedural \& Workflow Aids and Instructional Diagrams have relatively high proportions of professional / vocational subject ratings.

\begin{figure*}[tbp]
  \centering
  \includegraphics[width=0.58\textwidth]{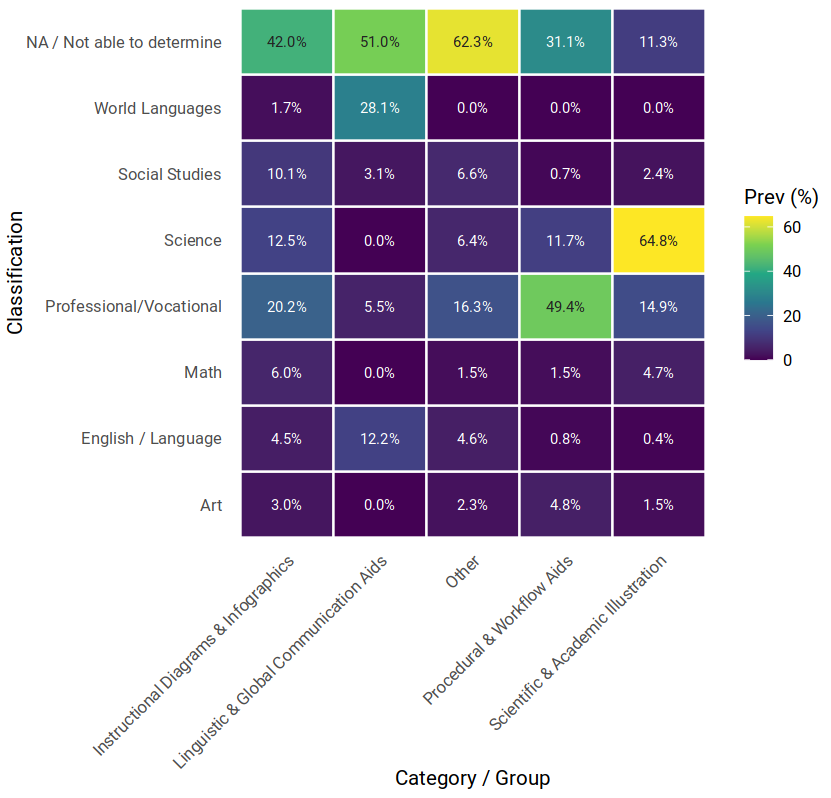}
  \caption{Prevalence of Subjects / Domains referenced in Educational \& Technical Information Media clusters.}
  \label{fig:edu_subjects}
\end{figure*}

Educational prompts were also rated on if the prompt referenced a particular grade or academic level---K--12, advanced or higher education, or vocational/professional. Many of these were NA, but of those rated there are some suggested patterns observed. In general there are a lot of K--12 level subject matter---especially the language learning aids. The workflow and procedural aids, and the scientific and academic illustrations have the highest level of advanced or higher ed grade levels.

\begin{figure*}[tbp]
  \centering
  \includegraphics[width=0.78\textwidth]{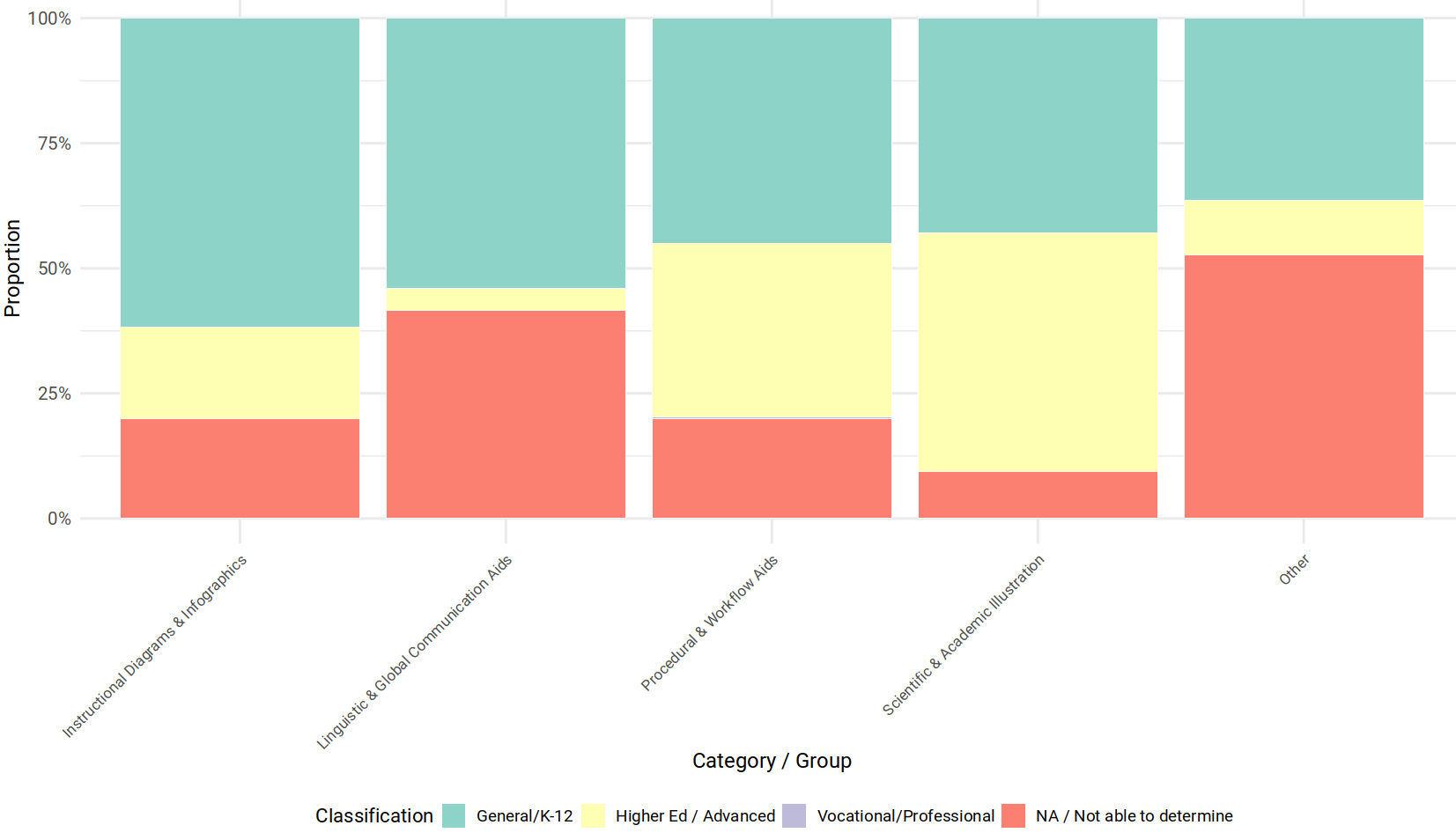}
  \caption{Prevalence of Grade / educational level referenced in Educational \& Technical Information Media clusters.}
  \label{fig:edu_grades}
\end{figure*}

\subsubsection*{Product Visualization \& E-Commerce Clusters: Prevalence of Features}

We classified conversations into Product Visualization and E-Commerce clusters focused on images visualizing products for listing and sales online, such as ``Footwear Product Visualization, Modification, and Display'', ``Apparel Visualization on Models and Mannequins'', and ``Cake Photography: Focus on Realistic Digital Staging.''

``Includes people,'' was perhaps surprisingly the highest prevalence in e-commerce (though many clusters described putting a specific product on a model, or showing items next to hands). After that, the most prevalent features here are compositional: Precise Positioning, Modify Clothing or Apparel, Modify Lighting, Negative Constraints.

\begin{figure}[htbp]
  \centering
  \includegraphics[width=\linewidth]{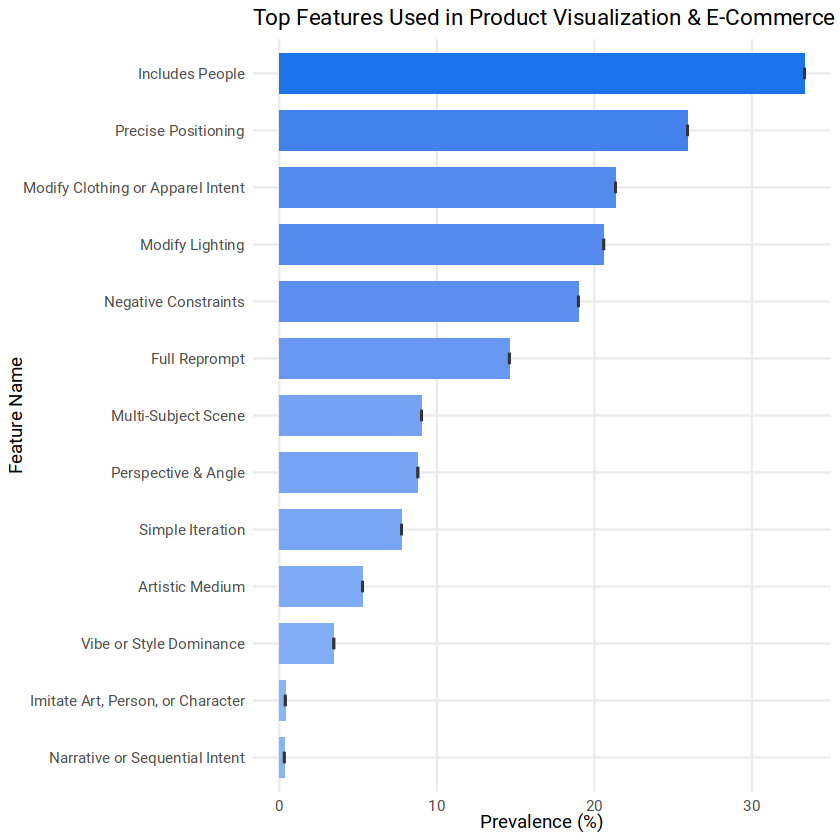}
  \caption{Prevalence of features in Product Visualization \& E-Commerce conversations.}
  \label{fig:features_ecommerce}
\end{figure}

\subsection{Analysis: Summary of Features Across Categories}
\label{sec:analysis_summary}

We compared the five high level categories of interest to each other across the binary features and multi-select features.

\subsubsection*{Multi-Select Features: Comparing Across Categories}

As discussed above, clusters of conversations were rated by a set of multiple choice features including text (text content prompts, text style prompts, no text / NA, or both content and style typography prompts); work context (small business / side hustle, professional / service based, or NA); stage of creation / ideation; and level of expertise demonstrated / required.

\begin{figure*}[tbp]
  \centering
  \includegraphics[width=0.65\textwidth]{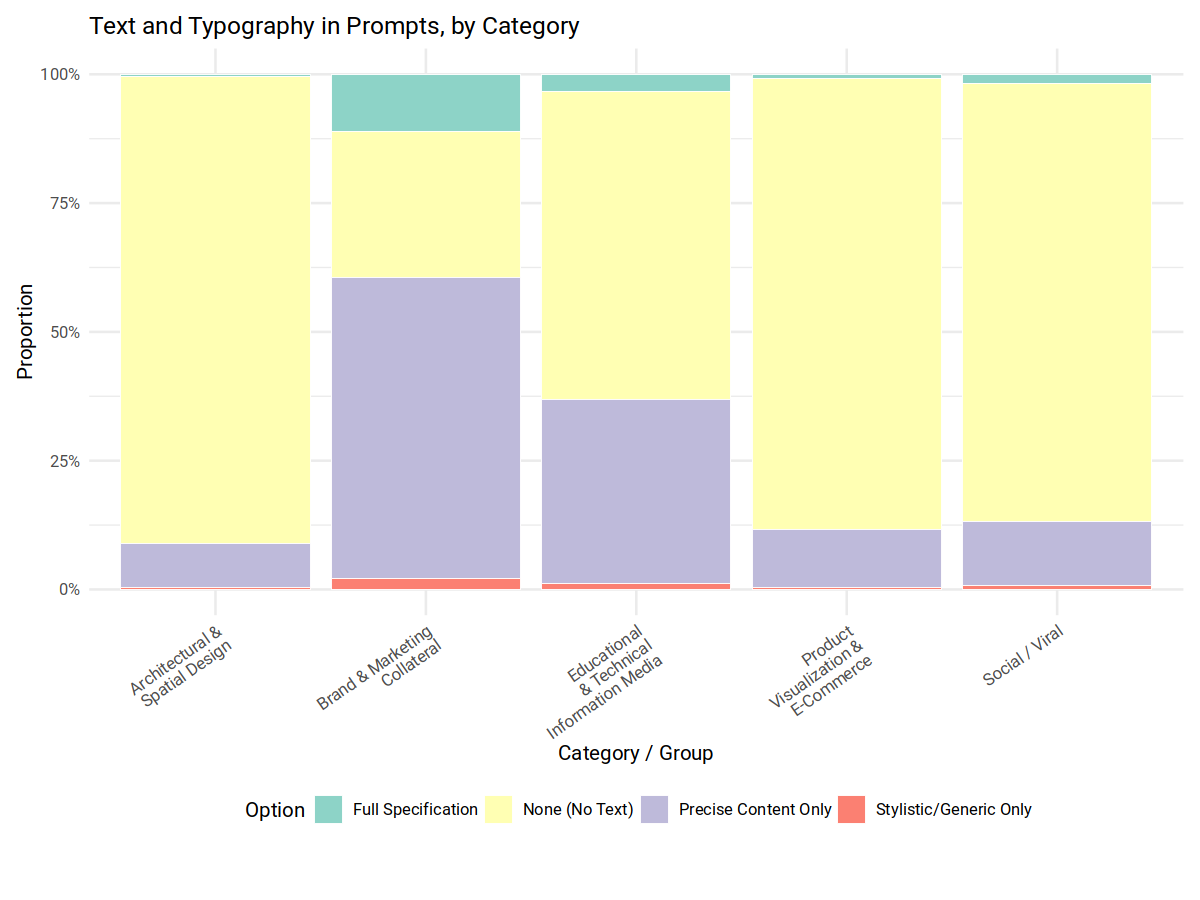}
  \caption{Prevalence of text and typography by high level conversation category.}
  \label{fig:multi_typography}
\end{figure*}

As shown in Figure~\ref{fig:multi_typography}, the vast majority of architectural, product visualization, and social / viral conversation clusters showed no text or typography in prompts. When there were typography instructions, it was frequently content-only. Educational \& Technical Information Media was more likely to include text or typography, around a third of the time. Brand \& Marketing Content had the most prevalent prompt requests for precise text and for full specifications.

\begin{figure*}[tbp]
  \centering
  \includegraphics[width=0.58\textwidth]{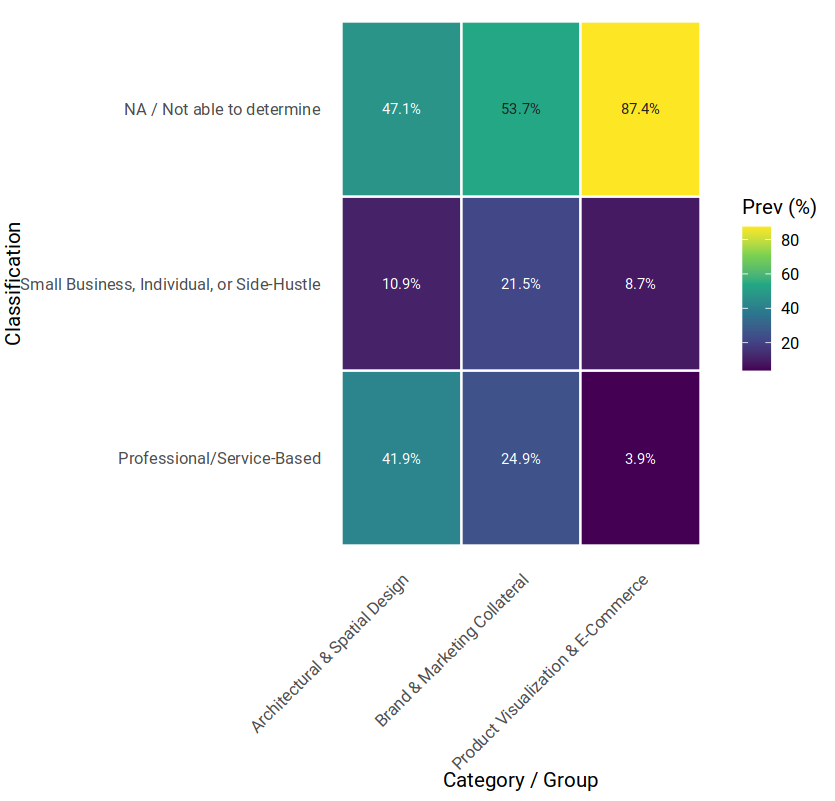}
  \caption{Prevalence of work role ratings by high level conversation category.}
  \label{fig:multi_work_role}
\end{figure*}

The majority of clusters were rated ``NA'' for work role (Product Visualization rated nearly 9 in 10 as NA). Architectural and Spatial Design had the highest prevalence of professional or service based ratings, and Brand \& Marketing Collateral had the highest prevalence of small business, individual, or side hustle ratings.

\begin{figure*}[tbp]
  \centering
  \includegraphics[width=0.58\textwidth]{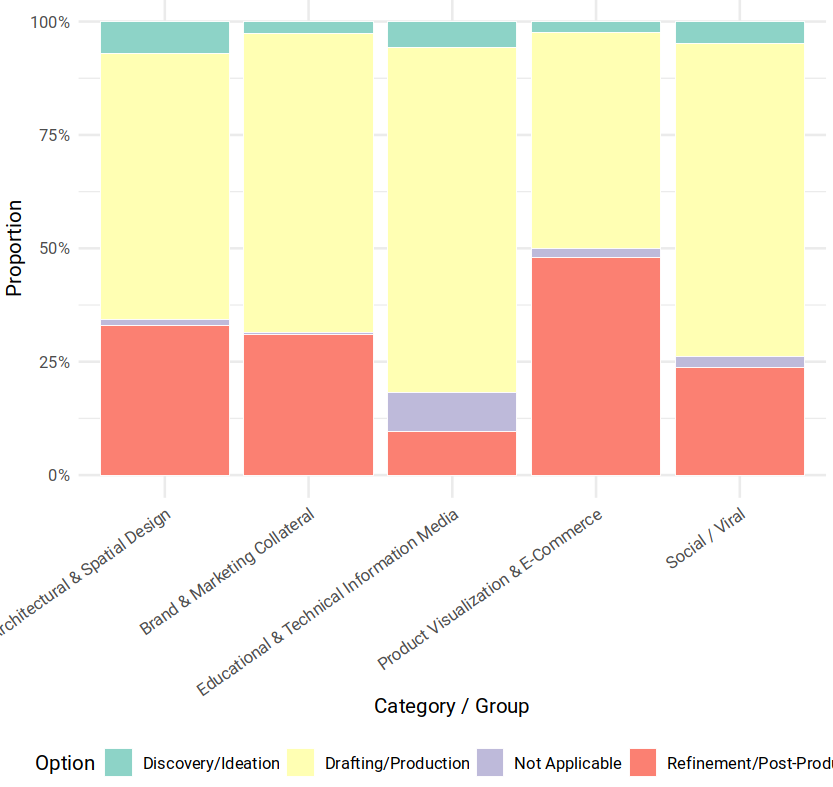}
  \caption{Prevalence of workflow phase ratings by high level conversation category.}
  \label{fig:multi_workflow_phase}
\end{figure*}

Nano Banana conversation clusters were also rated by workflow phase---Discovery / Ideation, Drafting / Production, and Refinement / Post-Production. For image generation and editing prompts, a plurality of conversations were rated as Drafting / Production, which is perhaps unsurprising. The next most prevalent was in refinement / post production---which was particularly high for product visualization, where prompts perhaps started with a product image and iterated to place the product in different contexts and angles.

\subsubsection*{Binary Feature Ratings: Summary Across High-Level Categories}
\begin{figure*}[tbp]
  \centering
  \includegraphics[width=1.0\textwidth]{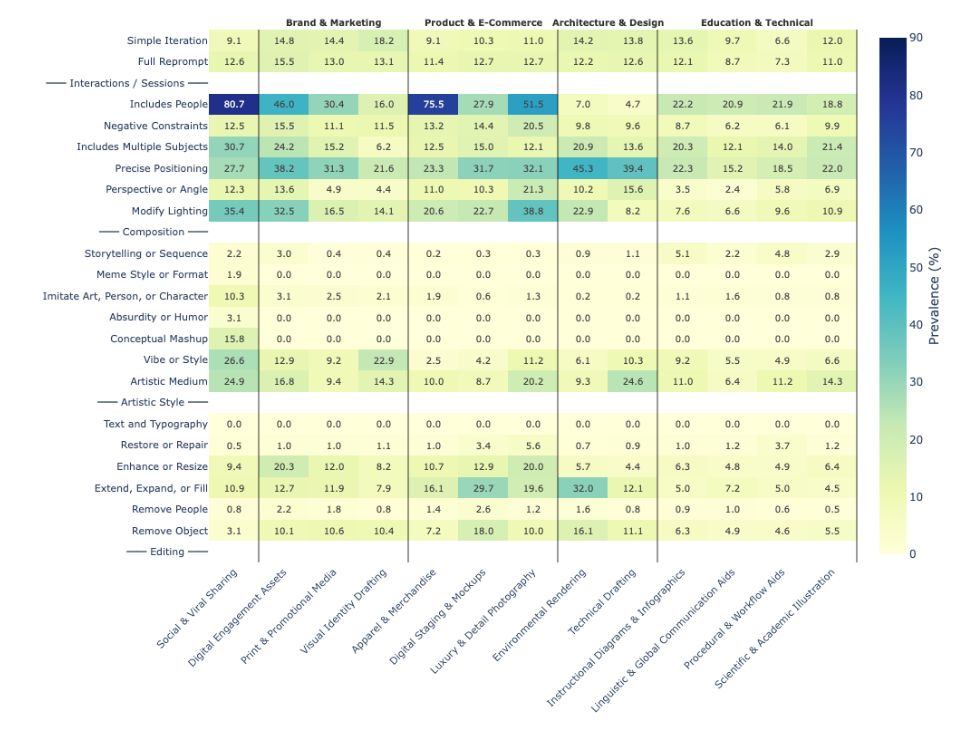}
  \caption{Summary heat map of binary feature ratings by high level conversation category.}
  \label{fig:binary_heatmap}
\end{figure*}

Figure~\ref{fig:binary_heatmap} summarizes binary feature prevalence across all five high-level categories and their subcategories.

\FloatBarrier

\subsection{Analysis: Countries with Highest Frequency of Features}
\label{sec:analysis_countries}

Social media in different languages and countries suggests that there are some differences in the way people are using image generation and editing tools (for example, \href{https://www.instagram.com/p/DOS0F4Piq3-/?hl=en}{cricket action figure posts} from India, and \href{https://www.instagram.com/p/DW1W0gXDL5D/}{anime soccer posts} from UAE). We sought to compare countries also on \emph{how} people used image generation and editing tools. We compared features by their prevalence in conversations from different countries (among countries with $>100$ conversations with those features).

Here we can observe some differences in the geographic distributions of the topics (people or not, fictional characters included) and styles of image generation and editing (lighting, meme formats, precise positioning), as well as how people interact with the tools during Nano Banana conversations (full re-prompting or not). These differences might be driven by country-level differences in social media trends, differences in user bases who have adopted Gemini and Nano Banana, seasonal events or holidays, or many other factors---here we simply report some differences we see to note that there is a geographic distribution of feature use.

\subsubsection*{Top countries by prevalence of features}
\label{sec:top_countries}

\begin{table}[htbp]
  \centering
  \caption{Top 10 countries by prevalence of precise positioning prompts.}
  \label{tab:countries_positioning}
  \footnotesize
  \begin{tabularx}{\columnwidth}{@{}r >{\raggedright\arraybackslash}X r@{}}
    \toprule
    \textbf{Rank} & \textbf{Country} & \textbf{Prevalence (\% Yes)} \\
    \midrule
     1 & Malaysia      & \pct{32.2}{35} \\
     2 & Singapore     & \pct{30.1}{35} \\
     3 & Estonia       & \pct{29.1}{35} \\
     4 & Finland       & \pct{29.0}{35} \\
     5 & Indonesia     & \pct{28.7}{35} \\
     6 & Taiwan        & \pct{28.6}{35} \\
     7 & Austria       & \pct{28.3}{35} \\
     8 & Thailand      & \pct{28.2}{35} \\
     9 & Germany       & \pct{28.2}{35} \\
    10 & United States & \pct{28.1}{35} \\
    \bottomrule
  \end{tabularx}
  \par\smallskip\noindent{\scriptsize\textit{Note:} Among countries with $>100$ conversations with the feature; bars scaled within table.\par}
\end{table}

\begin{table}[htbp]
  \centering
  \caption{Top 10 countries by prevalence of modifying lighting.}
  \label{tab:countries_lighting}
  \footnotesize
  \begin{tabularx}{\columnwidth}{@{}r >{\raggedright\arraybackslash}X r@{}}
    \toprule
    \textbf{Rank} & \textbf{Country} & \textbf{Prevalence (\% Yes)} \\
    \midrule
     1 & Somalia   & \pct{28.4}{35} \\
     2 & Malaysia  & \pct{28.3}{35} \\
     3 & India     & \pct{28.1}{35} \\
     4 & Nigeria   & \pct{27.7}{35} \\
     5 & Sri Lanka & \pct{27.5}{35} \\
     6 & Nepal     & \pct{24.2}{35} \\
     7 & Thailand  & \pct{24.1}{35} \\
     8 & Laos      & \pct{23.0}{35} \\
     9 & Pakistan  & \pct{22.9}{35} \\
    10 & Indonesia & \pct{22.4}{35} \\
    \bottomrule
  \end{tabularx}
  \par\smallskip\noindent{\scriptsize\textit{Note:} Among countries with $>100$ conversations with the feature; bars scaled within table.\par}
\end{table}

\begin{table}[htbp]
  \centering
  \caption{Top 10 countries by prevalence of imitating art, persons, or characters.}
  \label{tab:countries_imitating}
  \footnotesize
  \begin{tabularx}{\columnwidth}{@{}r >{\raggedright\arraybackslash}X r@{}}
    \toprule
    \textbf{Rank} & \textbf{Country} & \textbf{Prevalence (\% Yes)} \\
    \midrule
     1 & Chile         & \pct{5.2}{6} \\
     2 & Argentina     & \pct{5.0}{6} \\
     3 & United States & \pct{4.9}{6} \\
     4 & Australia     & \pct{4.6}{6} \\
     5 & Canada        & \pct{4.4}{6} \\
     6 & Mexico        & \pct{4.3}{6} \\
     7 & Japan         & \pct{4.1}{6} \\
     8 & Peru          & \pct{3.8}{6} \\
     9 & Guatemala     & \pct{3.8}{6} \\
    10 & Greece        & \pct{3.8}{6} \\
    \bottomrule
  \end{tabularx}
  \par\smallskip\noindent{\scriptsize\textit{Note:} Among countries with $>100$ conversations with the feature; bars scaled within table.\par}
\end{table}

\begin{table}[htbp]
  \centering
  \caption{Top 10 countries by prevalence of meme style or format prompts.}
  \label{tab:countries_meme}
  \footnotesize
  \begin{tabularx}{\columnwidth}{@{}r >{\raggedright\arraybackslash}X r@{}}
    \toprule
    \textbf{Rank} & \textbf{Country} & \textbf{Prevalence (\% Yes)} \\
    \midrule
     1 & Pakistan      & \pct{0.7}{0.8} \\
     2 & United States & \pct{0.7}{0.8} \\
     3 & Philippines   & \pct{0.6}{0.8} \\
     4 & Thailand      & \pct{0.6}{0.8} \\
     5 & India         & \pct{0.6}{0.8} \\
     6 & Viet Nam      & \pct{0.5}{0.8} \\
     7 & Mexico        & \pct{0.4}{0.8} \\
     8 & Indonesia     & \pct{0.4}{0.8} \\
     9 & Egypt         & \pct{0.4}{0.8} \\
    10 & Brazil        & \pct{0.4}{0.8} \\
    \bottomrule
  \end{tabularx}
  \par\smallskip\noindent{\scriptsize\textit{Note:} Among countries with $>100$ conversations with the feature; bars scaled within table.\par}
\end{table}
\begin{table}[htbp]
  \centering
  \caption{Top 10 countries by prevalence of absurdity or humor.}
  \label{tab:countries_absurdity}
  \footnotesize
  \begin{tabularx}{\columnwidth}{@{}r >{\raggedright\arraybackslash}X r@{}}
    \toprule
    \textbf{Rank} & \textbf{Country} & \textbf{Prevalence (\% Yes)} \\
    \midrule
     1 & Argentina     & \pct{2.6}{3} \\
     2 & Japan         & \pct{2.5}{3} \\
     3 & Poland        & \pct{2.5}{3} \\
     4 & Chile         & \pct{2.1}{3} \\
     5 & United States & \pct{2.1}{3} \\
     6 & Canada        & \pct{2.0}{3} \\
     7 & UK            & \pct{1.9}{3} \\
     8 & Netherlands   & \pct{1.8}{3} \\
     9 & Ukraine       & \pct{1.8}{3} \\
    10 & Colombia      & \pct{1.8}{3} \\
    \bottomrule
  \end{tabularx}
  \par\smallskip\noindent{\scriptsize\textit{Note:} Among countries with $>100$ conversations with the feature; bars scaled within table.\par}
\end{table}

\begin{table}[htbp]
  \centering
  \caption{Top 10 countries by prevalence of people in prompts.}
  \label{tab:countries_people}
  \footnotesize
  \begin{tabularx}{\columnwidth}{@{}r >{\raggedright\arraybackslash}X r@{}}
    \toprule
    \textbf{Rank} & \textbf{Country} & \textbf{Prevalence (\% Yes)} \\
    \midrule
     1 & India       & \pct{65.5}{70} \\
     2 & Iraq        & \pct{65.3}{70} \\
     3 & Somalia     & \pct{65.0}{70} \\
     4 & Nigeria     & \pct{63.5}{70} \\
     5 & Indonesia   & \pct{63.0}{70} \\
     6 & Malaysia    & \pct{62.6}{70} \\
     7 & Suriname    & \pct{62.1}{70} \\
     8 & Nepal       & \pct{61.8}{70} \\
     9 & Philippines & \pct{61.7}{70} \\
    10 & Congo       & \pct{60.8}{70} \\
    \bottomrule
  \end{tabularx}
  \par\smallskip\noindent{\scriptsize\textit{Note:} Among countries with $>100$ conversations with the feature; bars scaled within table.\par}
\end{table}

\begin{table}[htbp]
  \centering
  \caption{Top 10 countries by prevalence of clothing or apparel modification.}
  \label{tab:countries_clothing}
  \footnotesize
  \begin{tabularx}{\columnwidth}{@{}r >{\raggedright\arraybackslash}X r@{}}
    \toprule
    \textbf{Rank} & \textbf{Country} & \textbf{Prevalence (\% Yes)} \\
    \midrule
     1 & Bangladesh   & \pct{35.2}{40} \\
     2 & Morocco      & \pct{29.3}{40} \\
     3 & Egypt        & \pct{28.4}{40} \\
     4 & Sri Lanka    & \pct{27.9}{40} \\
     5 & Iraq         & \pct{27.7}{40} \\
     6 & Algeria      & \pct{26.4}{40} \\
     7 & India        & \pct{26.2}{40} \\
     8 & Saudi Arabia & \pct{26.0}{40} \\
     9 & Jordan       & \pct{24.5}{40} \\
    10 & Viet Nam     & \pct{22.8}{40} \\
    \bottomrule
  \end{tabularx}
  \par\smallskip\noindent{\scriptsize\textit{Note:} Among countries with $>100$ conversations with the feature; bars scaled within table.\par}
\end{table}

\begin{table}[htbp]
  \centering
  \caption{Top 10 countries by prevalence of full reprompts in conversations.}
  \label{tab:countries_reprompts}
  \footnotesize
  \begin{tabularx}{\columnwidth}{@{}r >{\raggedright\arraybackslash}X r@{}}
    \toprule
    \textbf{Rank} & \textbf{Country} & \textbf{Prevalence (\% Yes)} \\
    \midrule
     1 & Cambodia    & \pct{21.2}{25} \\
     2 & Hong Kong   & \pct{18.9}{25} \\
     3 & Singapore   & \pct{18.0}{25} \\
     4 & Philippines & \pct{17.9}{25} \\
     5 & Senegal     & \pct{17.8}{25} \\
     6 & Malaysia    & \pct{17.4}{25} \\
     7 & Indonesia   & \pct{17.4}{25} \\
     8 & Laos        & \pct{17.4}{25} \\
     9 & Sri Lanka   & \pct{17.0}{25} \\
    10 & Somalia     & \pct{16.7}{25} \\
    \bottomrule
  \end{tabularx}
  \par\smallskip\noindent{\scriptsize\textit{Note:} Among countries with $>100$ conversations with the feature; bars scaled within table.\par}
\end{table}

\FloatBarrier

\section{Discussion}
\label{sec:discussion}

We present the first large-scale, end-to-end privacy-preserving analysis of some of what people are using LLM AI Generation for, and how they are using it. While the work offers broad insights for application and use of image generation tools and use, we note key limitations:

\subsection{Limitations}
\label{sec:limitations}

\noindent\textbf{Reliance on prompt text to analyze intent.} Because we do not look at any images---reference images uploaded by users, or output images generated by Gemini---we rely only on summaries of text prompts and conversations put out by the Urania privacy-preserving tool. This means we may be missing the prevalence of some use cases, or missing some use cases entirely. For example, a user could upload a picture of their dog and the picture of a world leader---each picture including multiples of the features we seek to measure such as distinctive lighting and vibes, presence of objects or subjects or artistic styles, and write a prompt that says ``make this look like this,'' and Urania's analysis would have very limited accuracy on the users' intent in that conversation. The feature ratings measure if a user explicitly asked for a certain thing in the text, so those should be more accurate, but there is still some risk that we are miscounting by relying on text summaries only. Additionally, using prompt text summaries can measure what users \emph{ask for}, but it cannot analyze what Nano Banana provided back to users. These are measures of intents, not outputs (though iteration and re-prompt request statistics can shed some light on how satisfied users were with the outputs).

\noindent\textbf{Data sample.} We use a snapshot of data that may have seasonal or other effects. While large (2M conversations) it is only a small fraction of the total use.

\noindent\textbf{Reliability of LLM classification.} We use techniques that have been validated on public datasets, but the nature of the privacy preservation techniques from Urania mean that we cannot validate our classifications through human review of the individual prompts and conversations themselves. There is some uncertainty and likely noise in our results, which is why we see them as directional and to an extent ordinal, but not necessarily precise.

\noindent\textbf{Changing nature of tools and use cases.} This is a developing technology with changing users and a shifting user base, which reflects and is responding to changing trends in broader culture and social communications. This is a snapshot, a baseline set of readings from the first quarter of 2026, but is almost certainly not the final word on what people use LLM image generation and editing for, or how they will use it going forward.

\noindent\textbf{Lack of full context into user journeys.} These intent classifications, and in some cases features, are about narrow, cleanly definable aspects of conversations which include image generation prompts. Urania analyzes each conversation on its own, and only the conversations we sampled from Gemini usage. We cannot see if a user comes back and starts a different conversation on the same or a similar topic, nor can we see if a user shares an image or prompt external to Gemini, or pulls in a shared prompt or image from another source.

\subsection{Implications and future directions}
\label{sec:implications}

Acknowledging the above limitations, this analysis reveals several implications for how people are using Nano Banana image generation and editing.

\subsubsection*{Not just social media and viral trends}

While we do see some evidence of social media uses and viral trends, our analysis suggests that there are much broader and wider uses of image generation---in terms of subject matter and in terms of societal relevance. Even within social and viral trends there seem to be substantial differences in types of social and viral clusters and in features used between different countries, languages, and contexts.

Our analysis also shows clusters of educational and technical diagrams explaining details of real life objects and theoretical concepts, or offering step-by-step instructions for troubleshooting and repair. There are clusters around labeling objects with names and terms in multiple languages to help language learners. There are clusters around preparing products for sale on e-commerce sites, and those around creating brand assets and materials for businesses. And clusters that seem to be about both people considering redesigns of their own homes, gardens, and offices, and people preparing architectural or real estate designs for professional purposes. There is a wide variety of applications for these tools, and that diversity, along with overall growth of use of the tools, suggests that use cases are still emerging.

\subsubsection*{Complements or substitutes?}

We do not see, in the current analysis, a clear signal as to whether image generation and editing is complementary to or a substitute for alternative tools for imaging, or for creating marketing, e-commerce, architectural, educational, or social media content. One hypothesis shared in the press is that Nano Banana might be a replacement for image editing software \citep{langley2025}. But there is some evidence that is not entirely the case---for example one of the sub-clusters in Educational and Technical Media is around procedural and workflow aids for ``Image and Photo Editing Techniques and Instructions.'' From the Urania summary: ``This cluster focuses on detailed, step-by-step instructions and guides for image and photo editing \ldots\ like changing colors, removing backgrounds or objects, enhancing quality, resizing images, and applying effects using various software (Photoshop, Canva, etc.).'' At least in some cases, the AI LLM generation seems to be used to better understand and use \emph{existing image editing tools}.

\subsubsection*{Novelty and net new use cases}

While we see a lot of new use that didn't happen before Nano Banana was available in late 2025, we do not have clear signal from this analysis about whether and which types of use are novel and net new, vs.\ just significantly less expensive for users to generate. It seems likely that the use of reference images to generate a photo of you ``hugging your younger self'' is a novel one, made possible by the improved model capabilities to maintain a face or character across referenced and generated images. But it is worth further analysis, perhaps of prompts shared on social media, or different sources of data, to see if the shifts in model capabilities have resulted in newly created uses of the tools.

\subsubsection*{Learning curves and LLM tools}

This question is relevant to more than just image generation and editing tools, but as we analyze the specific features of use, or the specific ways in which people are prompting and using Nano Banana, we do not know how or why or how quickly they start to utilize more and different features. How do people become more or less savvy users of image generation and editing tools, and change what they are asking for and how? Our privacy preserved data is conversation-level only, and has no information about users or if they had more than one conversation. We cannot see anything at the individual level. But in future research, it might be worth experimenting to see if prompting more leads to greater iteration and inclusion of greater features---does using the tool and getting an image response lead people to ask more or different things? Do people learn from social media or shared prompts? Or do they not change their request behavior much at all?

\subsubsection*{Understanding broader impacts}

This analysis provides some view into how people are using image generation and editing tools and what they are using them for. Future research will have to look into the broader impacts and usefulness of these tools---and the images they create---in people's work, educational journeys, and social lives.

\section{Conclusion}
\label{sec:conclusion}

Nano Banana, launched in late 2025, is just over a year old, and updates to the model such as Nano Banana Pro and Nano Banana 2 are just months old. We are already seeing new (to the authors at least) playful and social ways of joking, simpler ways of restoring old family photos, of showing allegiance and fandom, of communicating with friends and family. Some argue we have even seen, through memes and image generation, new modes of diplomacy.

We see different (or at least less difficult to create) economic uses, from generating product visualizations for e-commerce, editing logos and brand collateral for small businesses and companies, to making visual mockups and architectural concepts, and infographics and step-by-step technical guides.

We are also seeing less commonly known or brand new things---using Nano Banana for conceptual explanation, to study new vocabulary, to develop quick prototypes of new products, or snapshots of what a new idea could look like.

New use cases and new tools will likely continue to appear in ways current users and developers have not anticipated, as people creatively work---and play---with new tools which are likely to continue to improve.

Our hope is by documenting the current state of some uses a half a year in, and the ``ingredients'' used to make these images, we can help spur further experimentation, innovation, and improvement from model makers and from the hundreds of millions of people who are using them---all while maintaining the privacy of individuals and their images.

\section{Acknowledgments}
\label{sec:acknowledgments}

We would like to thank colleagues who reviewed, ideated, and supported the work, including: RJ Chen, Jeff Naecker, Karina Livingston, Maya Shankar, Corinna Cortes, Patrick Thomas, Jane Park, Duncan Noble Smith, Lauren Winer, Jen Haslip, Alex Imas, Andrew Kim, Tanya Rodchenko, Preston McAfee, Janna Buckingham, Julia Wilkowski, and Rachel Hashimshoni.

\FloatBarrier
\bibliography{main}

\onecolumn
\appendix
\renewcommand{\thefigure}{A\arabic{figure}}
\setcounter{figure}{0}
\section{Appendix}
\label{sec:appendix}

\begin{figure}[!htbp]
  \centering
  \includegraphics[width=1.0\textwidth]{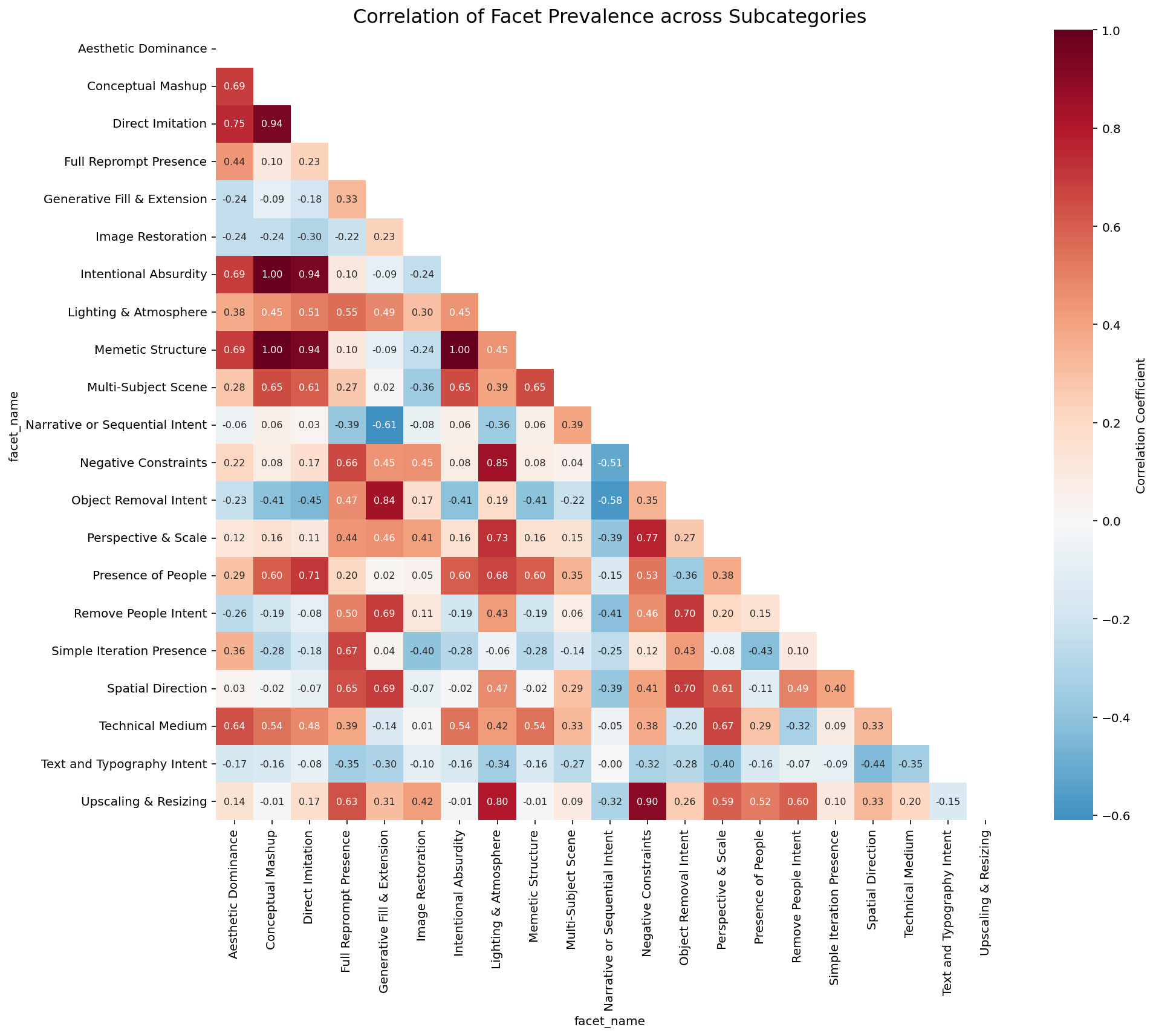}
  \caption{Correlation of feature prevalence across subcategories.}
  \label{fig:appendix_facet_correlation}
\end{figure}

\begin{figure}[!htbp]
  \centering
  \includegraphics[width=0.95\textwidth]{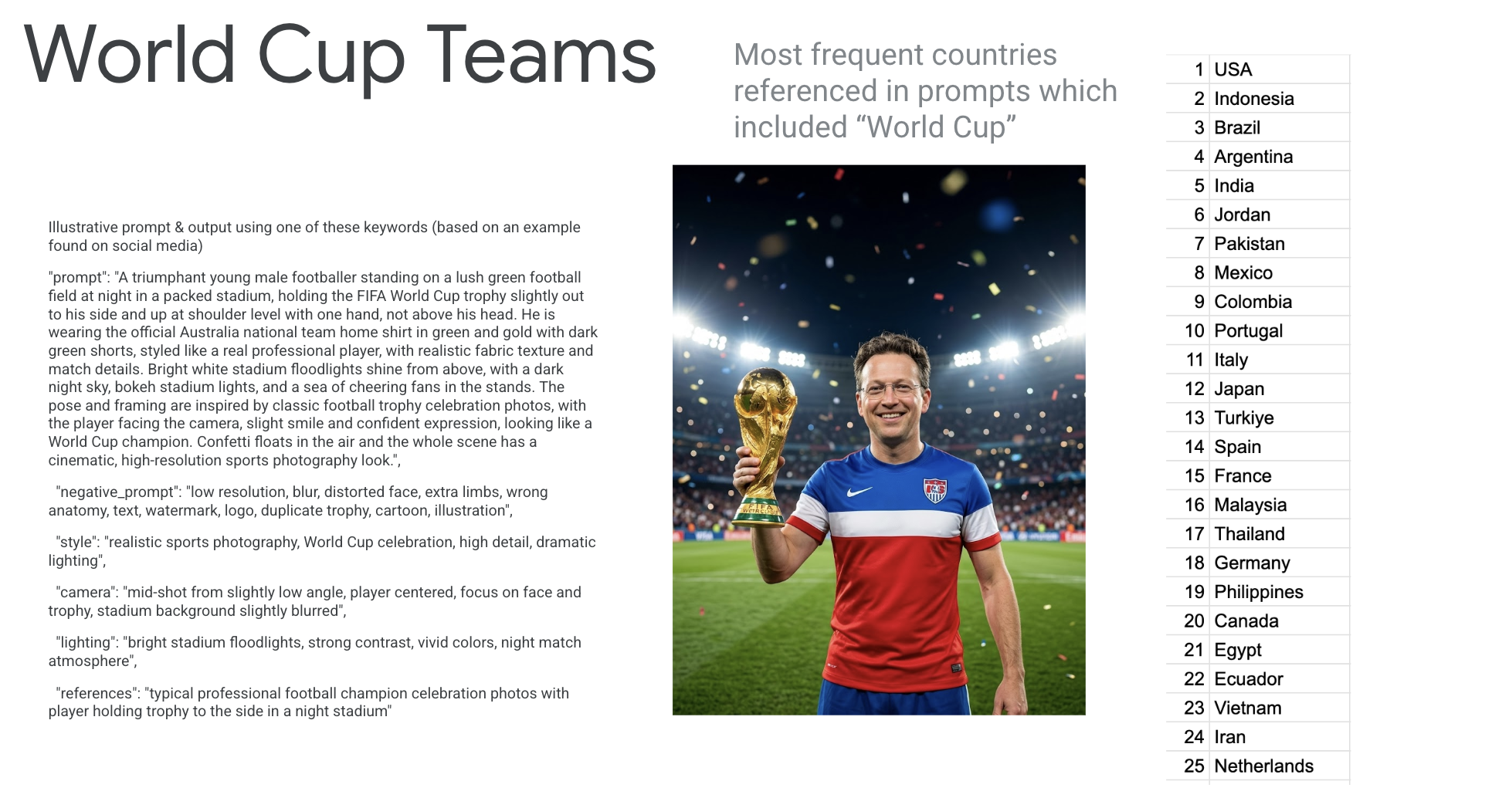}
  \caption{Illustrative ``World Cup'' social media prompt, generated output, and most frequently referenced countries (image of co-author included with permission).}
  \label{fig:appendix_world_cup}
\end{figure}

\begin{figure}[!htbp]
  \centering
  \includegraphics[width=0.95\textwidth]{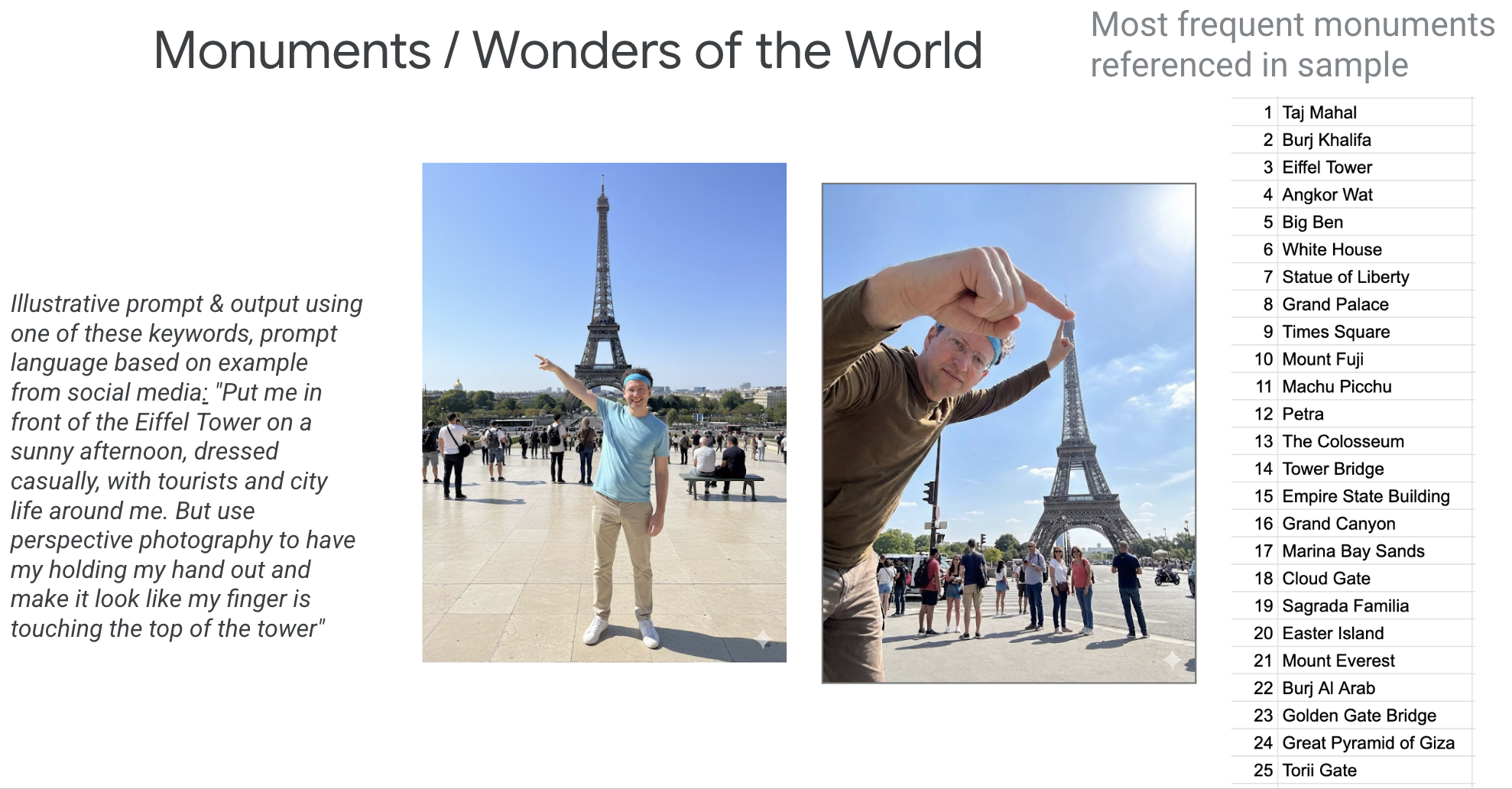}
  \caption{Illustrative monuments / wonders of the world prompt, generated outputs, and most frequently referenced monuments (images of coauthors included with permissions).}
  \label{fig:appendix_monuments}
\end{figure}

\end{document}